\documentstyle[11pt,psfig,epsfig,subfigure]{article}
\begin{document}
\renewcommand{\baselinestretch}{1.5}
%\vspace*{-0.5in}
%\begin{flushright}
%CDF/ANAL/ELECTROWEAK/CDFR/6233  \\
%PRD draft, Version 2.0\\ 
%\today\\
%\end{flushright}
%\vspace{1.0in}
\Large
\begin{center}
{\boldmath \bf  Measurement of the polar-angle distribution of leptons from $W$ boson decay as a function of the $W$ transverse momentum in $p\bar{p}$ collisions at $\sqrt{s}$=1.8 TeV}
\end{center}
\normalsize
\font\eightit=cmti8
\def\r#1{\ignorespaces $^{#1}$}
\hfilneg
\begin{sloppypar}
\noindent
D.~Acosta,\r {14} T.~Affolder,\r 7 M.G.~Albrow,\r {13} D.~Ambrose,\r {36}   
D.~Amidei,\r {27} K.~Anikeev,\r {26} J.~Antos,\r 1 
G.~Apollinari,\r {13} T.~Arisawa,\r {50} A.~Artikov,\r {11}
% ...
\end{sloppypar}
\vskip .026in
\begin{center}
(CDF Collaboration)
\end{center}

\vskip .026in
\begin{center}
\r 1  {\eightit Institute of Physics, Academia Sinica, Taipei, Taiwan 11529, 
Republic of China} \\
\r 2  {\eightit Argonne National Laboratory, Argonne, Illinois 60439} \\
\r 3  {\eightit Istituto Nazionale di Fisica Nucleare, University of Bologna,
I-40127 Bologna, Italy} \\
\r 4  {\eightit Brandeis University, Waltham, Massachusetts 02254} \\
\r 5  {\eightit University of California at Davis, Davis, California  95616} \\
\r 6  {\eightit University of California at Los Angeles, Los 
Angeles, California  90024} \\ 
\r 7  {\eightit University of California at Santa Barbara, Santa Barbara, California 
93106} \\ 
\r 8 {\eightit Instituto de Fisica de Cantabria, CSIC-University of Cantabria, 
39005 Santander, Spain} \\
\r 9  {\eightit Carnegie Mellon University, Pittsburgh, Pennsylvania  15213} \\
\r {10} {\eightit Enrico Fermi Institute, University of Chicago, Chicago, 
Illinois 60637} \\
\r {11}  {\eightit Joint Institute for Nuclear Research, RU-141980 Dubna, Russia}
\\
\r {12} {\eightit Duke University, Durham, North Carolina  27708} \\
\r {13} {\eightit Fermi National Accelerator Laboratory, Batavia, Illinois 
60510} \\
\r {14} {\eightit University of Florida, Gainesville, Florida  32611} \\
\r {15} {\eightit Laboratori Nazionali di Frascati, Istituto Nazionale di Fisica
               Nucleare, I-00044 Frascati, Italy} \\
\r {16} {\eightit University of Geneva, CH-1211 Geneva 4, Switzerland} \\
\r {17} {\eightit Glasgow University, Glasgow G12 8QQ, United Kingdom}\\
\r {18} {\eightit Harvard University, Cambridge, Massachusetts 02138} \\
\r {19} {\eightit Hiroshima University, Higashi-Hiroshima 724, Japan} \\
\r {20} {\eightit University of Illinois, Urbana, Illinois 61801} \\
\r {21} {\eightit The Johns Hopkins University, Baltimore, Maryland 21218} \\
\r {22} {\eightit Institut f\"{u}r Experimentelle Kernphysik, 
Universit\"{a}t Karlsruhe, 76128 Karlsruhe, Germany} \\
\r {23} {\eightit Center for High Energy Physics: Kyungpook National
University, Taegu 702-701; Seoul National University, Seoul 151-742; and
SungKyunKwan University, Suwon 440-746; Korea} \\
\r {24} {\eightit Ernest Orlando Lawrence Berkeley National Laboratory, 
Berkeley, California 94720} \\
\r {25} {\eightit University College London, London WC1E 6BT, United Kingdom} \\
\r {26} {\eightit Massachusetts Institute of Technology, Cambridge,
Massachusetts  02139} \\   
\r {27} {\eightit University of Michigan, Ann Arbor, Michigan 48109} \\
\r {28} {\eightit Michigan State University, East Lansing, Michigan  48824} \\
\r {29} {\eightit Institution for Theoretical and Experimental Physics, ITEP,
Moscow 117259, Russia} \\
\r {30} {\eightit University of New Mexico, Albuquerque, New Mexico 87131} \\
\r {31} {\eightit Northwestern University, Evanston, Illinois  60208} \\
\r {32} {\eightit The Ohio State University, Columbus, Ohio  43210} \\
\r {33} {\eightit Osaka City University, Osaka 588, Japan} \\
\r {34} {\eightit University of Oxford, Oxford OX1 3RH, United Kingdom} \\
\r {35} {\eightit Universita di Padova, Istituto Nazionale di Fisica 
          Nucleare, Sezione di Padova, I-35131 Padova, Italy} \\
\r {36} {\eightit University of Pennsylvania, Philadelphia, 
        Pennsylvania 19104} \\   
\r {37} {\eightit Istituto Nazionale di Fisica Nucleare, University and Scuola
               Normale Superiore of Pisa, I-56100 Pisa, Italy} \\
\r {38} {\eightit University of Pittsburgh, Pittsburgh, Pennsylvania 15260} \\
\r {39} {\eightit Purdue University, West Lafayette, Indiana 47907} \\
\r {40} {\eightit University of Rochester, Rochester, New York 14627} \\
\r {41} {\eightit Rockefeller University, New York, New York 10021} \\
\r {42} {\eightit Instituto Nazionale de Fisica Nucleare, Sezione di Roma,
University di Roma I, ``La Sapienza," I-00185 Roma, Italy}\\
\r {43} {\eightit Rutgers University, Piscataway, New Jersey 08855} \\
\r {44} {\eightit Texas A\&M University, College Station, Texas 77843} \\
\r {45} {\eightit Texas Tech University, Lubbock, Texas 79409} \\
\r {46} {\eightit Institute of Particle Physics, University of Toronto, Toronto
M5S 1A7, Canada} \\
\r {47} {\eightit Istituto Nazionale di Fisica Nucleare, University of Trieste/\
Udine, Italy} \\
\r {48} {\eightit University of Tsukuba, Tsukuba, Ibaraki 305, Japan} \\
\r {49} {\eightit Tufts University, Medford, Massachusetts 02155} \\
\r {50} {\eightit Waseda University, Tokyo 169, Japan} \\
\r {51} {\eightit University of Wisconsin, Madison, Wisconsin 53706} \\
\r {52} {\eightit Yale University, New Haven, Connecticut 06520} \\
\r {53} {\eightit CCLRC Rutherford Appleton Laboratory, Didcot OX11 0QX, UK} \\
\end{center}

\renewcommand{\baselinestretch}{1.5}
\normalsize

\begin{center}
{\bf Abstract}
\end{center}
We present a measurement of the coefficient $\alpha_2$ of the leptonic polar-angle distribution from $W$ boson decays, as a function of the $W$ transverse momentum. The measurement uses an 80$\pm$4 pb$^{-1}$ sample of $p\bar{p}$ collisions at $\sqrt{s}$=1.8 TeV collected by the CDF detector and includes data from both the $W\rightarrow e+\nu$ and $W\rightarrow\mu+\nu$ decay channels. We fit the $W$ boson transverse mass distribution to a set of templates from a Monte Carlo event generator and detector simulation in several ranges of the $W$ transverse momentum. The measurement agrees with the Standard Model expectation, whereby the ratio of longitudinally to transversely polarized $W$ bosons, in the Collins-Soper $W$ rest frame, increases with the $W$ transverse momentum at a rate of approximately 15\% per 10 GeV/$c$.\\
\\
PACS numbers: 12.15.Ji, 12.38.Qk, 13.85.Qk, 13.38.Be, 14.70.Fm
\vspace*{0.5in}
\clearpage

\normalsize
\begin{center}{\bf I. INTRODUCTION}\end{center}
\pagestyle{plain}
\pagenumbering{arabic}    % roman numeral pagenumbers

\begin{figure}[h]
\begin{center}
\epsfig{file=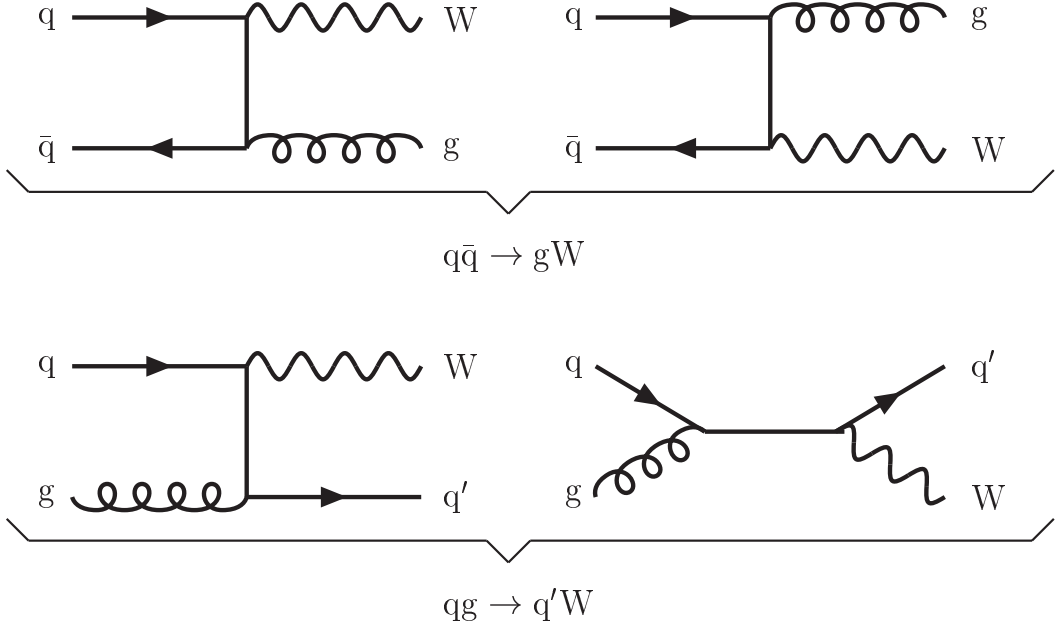,width=14cm,angle=0} 
\end{center}
\caption{\small{The QCD leading-order processes that give rise to $W$ production at high-$p_T^W$. In the top diagrams a gluon is radiated from one of the scattering quarks. In the bottom diagrams a quark-gluon scattering produces a $W$, together with a quark.}}
\label{fig:loqqbar}
\end{figure}

\begin{figure}[h]
\begin{center}
\epsfig{file=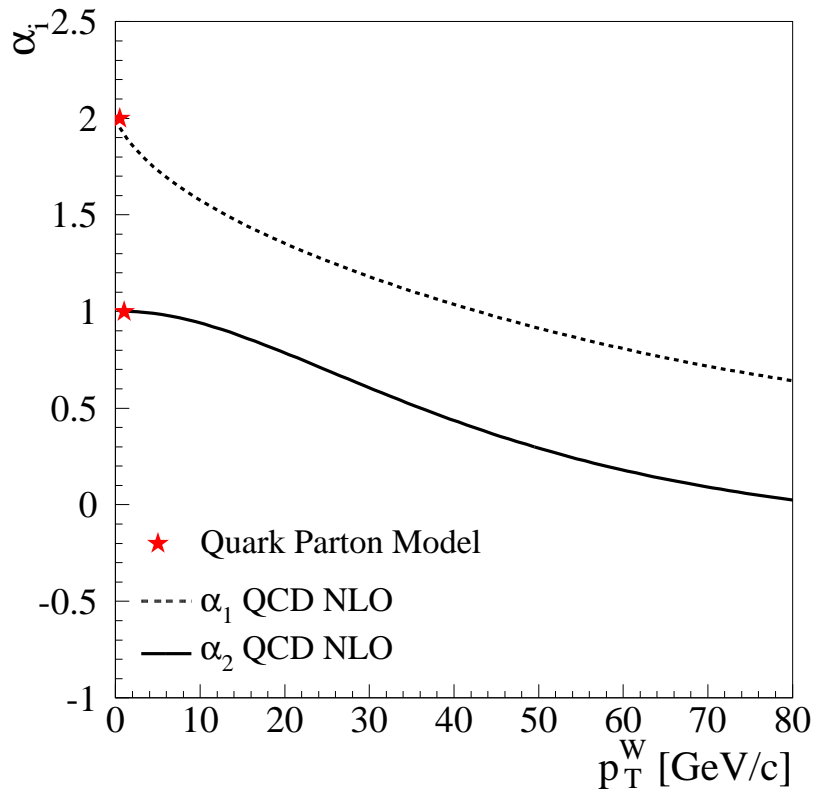,width=12cm,angle=0} 
\end{center}
\vspace{-0.5cm}
\caption{\small{Theoretical NLO-QCD calculation of $\alpha_2$ and $\alpha_1$ vs. $p_T^W$. The limit $p_T^W\rightarrow 0$ GeV/$c$ is the Quark Parton Model, for which $\alpha_2=1$ and $\alpha_1=2$ .}}
\label{fig:theo_pred}
\end{figure}

According to the Standard Model (SM), the polarization of $W$ bosons produced at high transverse momentum ($p_T^W$) is strongly affected by initial-state gluon radiation and quark-gluon scattering (the QCD leading-order diagrams for high-$p_T$ $W$ production are shown in Fig.~\ref{fig:loqqbar}). The angular distribution of the leptons from the $W\rightarrow \ell+\nu$ decay reflects the changes in the $W$ polarization. In the Collins-Soper $W$ rest frame \cite{footnote} the dependence of the cross section on the leptonic polar-angle at hadron level can be parametrized as
\begin{equation}\frac{{\rm d}\sigma}{{\rm d}\cos\theta_{CS}} \propto (1-Q \alpha_1 \cos\theta_{CS} +\alpha_2 \cos^2\theta_{CS}),
\label{equ:qcddist}
\end{equation}
where $Q$ is the lepton charge. The effects of QCD contribute to the coefficients $\alpha_1$ and $\alpha_2$, which are functions of $p_T^W$. Fig.~\ref{fig:theo_pred} shows the theoretical expectation for $\alpha_1(p_T^W)$ and $\alpha_2(p_T^W)$, neglecting a correction from sea-quarks, calculated up to next-to-leading order in QCD \cite{mirkes,d0paper}. Sea quarks give an opposite sign contribution to the $\cos\theta_{CS}$ term when the $W$ is produced by an antiquark in the proton and a quark in the antiproton, reducing the value of $\alpha_1$. Only in the limit $p_T^W\rightarrow 0$ GeV/$c$, when $\alpha_1=2$ and $\alpha_2=1$, does Eq.~(\ref{equ:qcddist}) describe the distribution of leptons from a transversely polarized $W$ boson: ${\rm d}\sigma/{\rm d}\cos\theta_{CS}\propto(1-Q\cos\theta_{CS})^2$, which is typical of a pure $V-A$ interaction. As $\alpha_2$ decreases, the contribution from longitudinally polarized $W$ bosons increases and so does the probability for the decay lepton to be emitted at large polar angle. On the other hand, $\alpha_1$ measures the forward-backward leptonic-decay asymmetry. Fig.~\ref{fig:theo_pred} indicates that the asymmetry is reduced at higher $p_T^W$.

Understanding how QCD corrections affect lepton angular distributions is important in the measurement of the $W$ mass ($M_W$) and rapidity distributions in $p\bar{p}$ experiments. The lepton angular distribution changes the shape of the transverse mass distribution, which is used to measure $M_W$. It has been estimated that an uncertainty of $\pm$1\% on $\alpha_2$ corresponds to a shift of the measured value of $M_W$, determined by fitting the transverse mass distribution, of approximately $\pm$10 MeV/$c^2$ \cite{mythesis}.

We present the measurement of $\alpha_2$ at various $W$ transverse momenta, using both the electron and muon channels. The sensitivity for a measurement of $\alpha_1$ is too low, due to the fact that the sign of $\cos\theta_{CS}$ is undetermined without a full reconstruction of the kinematics of the neutrino from the $W$ decay. Hence, the only sensitivity to $\alpha_1$ comes from the correlation between the geometrical acceptance of the detector and the phase space of the observed events. The current best measurement of $\alpha_2$ is reported in \cite{d0paper}. The results presented here reduce the uncertainty on $\alpha_2$ by about 50\% up to $p_T^W\sim$ 30 GeV/$c$, and are of comparable uncertainty at higher transverse momenta of the $W$.

For completeness, the cross section differential in the azimuthal and polar lepton angles can be expressed in the most general form as
\begin{eqnarray}
\frac{{\rm d}^4\sigma}{{\rm d} (p_T^W)^2~{\rm d}y~{\rm dcos}\theta_{CS}~{\rm d}\phi_{CS}} & = & \frac{3}{16\pi}\frac{{\rm d}^2\sigma^{TOT}}{{\rm d}(p_T^W)^2 {\rm d}y} [(1+\cos^2\theta_{CS})+ \nonumber \\
 & & +\frac{1}{2}A_0(1-3\cos^2\theta_{CS})-Q A_1 \sin 2\theta_{CS} \cos\phi_{CS}+ \nonumber \\
 & & +\frac{1}{2} A_2 \sin^2\theta_{CS} \cos2\phi_{CS}+A_3 \sin\theta_{CS} \cos\phi_{CS}+ \nonumber \\
 & & -Q A_4 \cos\theta_{CS} +A_5 \sin^2\theta_{CS} \sin 2\phi_{CS}+ \nonumber \\
 & & -Q A_6 \sin 2\theta_{CS} \sin\phi_{CS}+A_7 \sin\theta_{CS} \sin\phi_{CS}], 
\label{equ:qcdangular}
\end{eqnarray}
where $y$ is the rapidity of the $W$ boson, $\sigma^{TOT}$ is the total (angle integrated) rate, and the $A_i$ terms weight the relative contributions to the total cross section due to the different polarizations of the $W$ boson. By integrating Eq.~(\ref{equ:qcdangular}) over $\phi$ and comparing with Eq.~(\ref{equ:qcddist}) it follows that
\begin{equation}
\alpha_1 =\frac{2A_4}{2+A_0}, \quad
\alpha_2=\frac{2-3A_0}{2+A_0},
\end{equation}
which relates the $\alpha_1$ and $\alpha_2$ with the $A_i$ coefficients. The $A_i$ coefficients are explicitly calculated in \cite{mirkes,jstrolog}.

This paper is structured as follows: Sections II and III describe the CDF detector and the $W$ boson data sample, Sections IV and V outline the measurement method and detail the Monte Carlo event generator and detector simulation. Section VI contains the estimate of the background to the $W$ data sample and Section VII summarizes the fits and the systematic uncertainties. The results and conclusions are presented in Section VIII.\\

\begin{figure}[h]
\begin{center}
\psfig{file=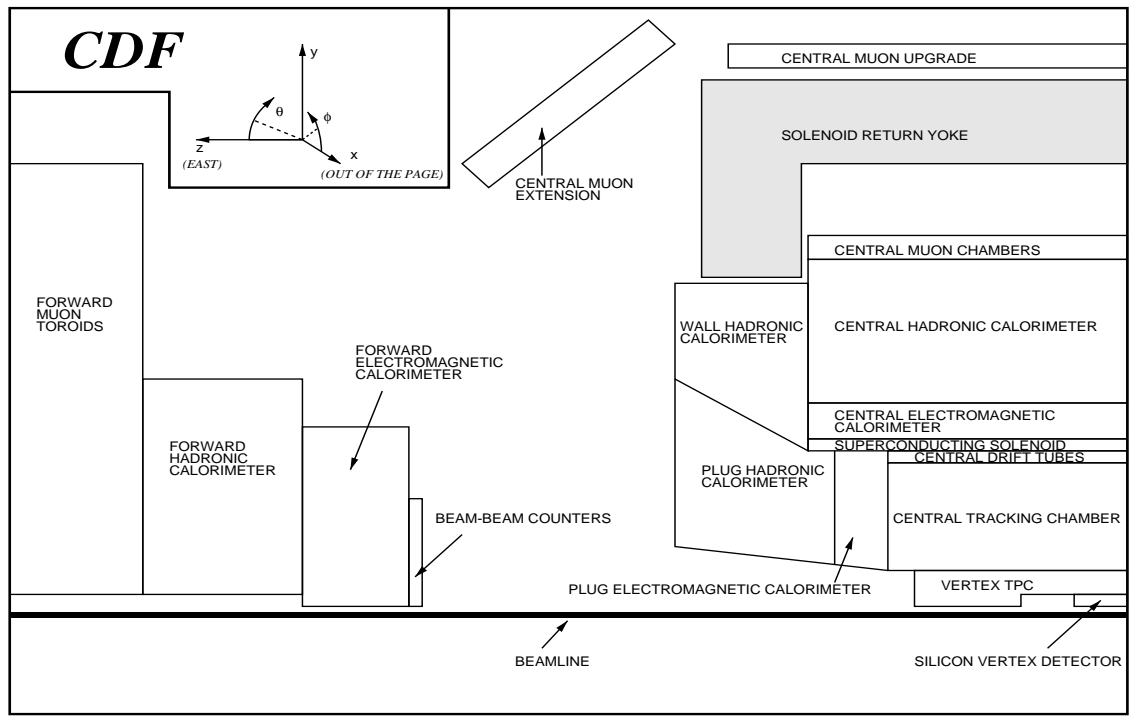,width=15cm,angle=0} 
\end{center}
\caption{\small{One quarter of the CDF detector. The detector is symmetric about the interaction point. This is the configuration for Run Ib.}}
\label{fig:cdfdetector}
\end{figure}

\begin{center}{\bf II. THE COLLIDER DETECTOR AT FERMILAB (CDF)}\end{center}
A complete description of the CDF detector can be found elsewhere \cite{cdf}. We describe here only the components relevant to this work. CDF uses a cylindrical coordinate system ($r$, $\phi$, $z$) with the origin at the center of the detector and the $z$-axis along the nominal direction of the proton beam. We define the polar angle $\theta$ as the angle measured with respect to the $z$-axis and the pseudo-rapidity ($\eta$) as $\eta=-\ln(\tan(\theta/2))$. A schematic drawing of one quadrant of the CDF detector is shown in Fig.~\ref{fig:cdfdetector}.\\

\begin{center}{\bf A: Tracking}\end{center}
The CDF tracking system in Run I consists of three tracking detectors: a silicon vertex detector (SVX$^\prime$), a vertex time projection chamber (VTX) and an open-cell multiwire drift chamber (CTC). The tracking system is immersed in a 1.4 T solenoidal magnetic field aligned with the $z$-axis. The silicon vertex detector \cite{svx} is a silicon microstrip detector that covers a region in radius from 2.86 to 7.87 cm. It is divided into two identical ``barrels'' which surround the beampipe on opposite sides of the $z$ = 0 plane. Each barrel consists of four radial layers of silicon strip detectors, and each layer is divided in azimuth into 30$^\circ$ wedges. The microstrips run parallel to the $z$ direction so that the SVX$^\prime$ tracks particles in $r-\phi$. The VTX \cite{vtx} is a set of 28 time projection chambers, each 9.4 cm in length, surrounding the SVX$^\prime$ detector. It provides the $z$ position of the interaction point with a resolution of 1 to 2 mm. The CTC \cite{ctc}, which extends out to a radius of 138 cm and $|z|<$ 160 cm, measures a three-dimensional track by providing up to 60 axial and 24 stereo position measurements. The basic drift cell has a line of 12 sense wires strung parallel to the $z$-axis for axial measurements or 6 sense wires tilted $\pm$3$^\circ$ in $\phi$ for stereo measurements. The set of all drift cells located at the same radius from the origin of the detector is called a super-layer.

In this analysis the CTC is used for the tracking and VTX and SVX$^\prime$ are only used to provide vertex information. The CTC track is constrained to point to the event vertex. The $z$ location of the vertex is determined with the VTX, and the position in $r-\phi$ is determined from the beam line measured with the SVX$^\prime$. The result of this procedure is a significant improvement in the CTC resolution. The momentum resolution of such tracks is $\sigma(p_T)/p_T = [(0.0009~p_T)^2+(0.0066)^2]^{1/2}$ with $p_T$ measured in units of GeV/$c$.\\

\begin{center}{\bf B:  Calorimetry}\end{center}
The CDF calorimetry is provided by four different calorimeter systems with a nearly contiguous coverage out to $|\eta|$ = 4.2$~$. Three of the four systems have both electromagnetic (EM) and hadronic (HA) calorimetry.  They are called ``Central'' (CEM, CHA), ``Wall'' (WHA), ``Plug'' (PEM, PHA) and ``Forward'' (FEM, FHA). The central and wall calorimeters are scintillator based, whereas the plug and forward calorimeters are a sandwich of proportional tube arrays with lead (PEM) or steel (PHA) absorber, and they are all segmented into towers which point back to the nominal interaction point.

The CEM \cite{cem} provides electron and photon energy measurements in the region $|\eta|<$ 1.1 with resolution $\sigma_E/E = 13.5\%/\sqrt{E\sin\theta} \oplus 1.5\%$, where $E$ is measured in units of GeV and $\oplus$ indicates sum in quadrature. The CEM is physically separated into two halves, one covering $\eta>0$  and one covering $\eta<0$. Both halves are divided in azimuth into 24 wedges that subtend 15$^\circ$ each. Each wedge extends along the $z$-axis  for 246 cm and is divided into ten projective towers of approximately 0.1 units in $\eta$. 
The CEM  is 18 radiation lengths thick and consists of 31 layers of plastic scintillator interleaved with 30 layers of lead sheets. A proportional chamber (CES) measures the electron shower position in the $\phi$ and {\it z} directions at a depth of $\sim$ 6 radiation lengths in the CEM. The CES module in each wedge is a multi-wire proportional chamber with 64 anode wires oriented parallel to the beam axis. The cathodes are segmented into 128 strips perpendicular to anode wires. An electron and photon shower typically spans several CES channels in each dimension. When CTC tracks made by electrons from $W$ boson decays are extrapolated to the CES ($r\approx$ 184 cm), the CTC extrapolation and the CES shower position match to 0.22 cm (rms) in azimuth and 0.46 cm (rms) in $z$. Both CES/CTC position matching and the CES shower shape are used as electron identification variables.

The PEM provides energy measurement in the range 1.1$<|\eta|<$2.4 and the FEM covers 2.2$<|\eta|<$4.2. The towers subtend approximately 0.1 in pseudorapidity by 5$^\circ$ in $\phi$. Details of the plug and forward calorimeters can be found in \cite{plug,forward}.

All the calorimeters are used to measure missing transverse energy and the central electromagnetic calorimeter provides the energy measurement for the electrons in this analysis.\\ 

\begin{center}{\bf C: Muon Systems}\end{center}
Three systems of scintillators and proportional chambers are used to identify muons in this analysis. A four-layer array of drift chambers, embedded in each wedge directly outside of the CHA, form the central muon detection system (CMU) \cite{cmu1,cmu2}. The CMU covers the region $|\eta|<$ 0.6 and measures a four-point trajectory (called a ``stub'') with an accuracy of 250 $\mu$m per point in $r-\phi$. Charge division gives an accuracy of 1.2 mm per point in $z$. A 0.6 m-thick layer of steel separates the CMU from a second four-layer array of drift chambers (CMP). Requiring a muon to have a stub in the CMP reduces the background due to hadrons and in-flight decays by approximately a factor of ten. The CMU covers approximately 84\% of the solid angle for $|\eta|<$ 0.6, while 63\% is covered by the CMP, and 53\% by both. Additional four-layer muon chambers (CMX) with partial (70\%) azimuthal coverage lie within 0.6 $<|\eta|<$ 1.0.\\

\begin{center}{\bf D: Trigger Requirements}\end{center}
The CDF trigger \cite{trigger} is a three-level system that selects events for recording to magnetic tape. The first two levels of the trigger consist of dedicated electronics. At Level$~$1, electrons are selected by requiring the presence of deposited energy above 8 GeV in a trigger tower (one trigger tower is two physical towers, with a width in pseudorapidity of $\Delta\eta$=0.2). Muons are selected by requiring the presence of a track-stub in the CMU or CMX and, where there is coverage, a track-stub in the CMP in coincidence with the CMU. The Level$~$2 trigger starts after a Level 1 trigger has accepted an event. Trigger towers in the calorimeters are combined into clusters of total or electromagnetic energy by a hardware cluster finder. Clusters and stubs are then matched to tracks found in the CTC by the fast hardware tracking processor. The third-level trigger uses software based on optimized offline reconstruction code to analyze the whole event.

\begin{center}{\bf III. DATA SELECTION}\end{center}
The data presented here were collected by the CDF detector at the Tevatron collider between 1994 and 1995  (``Run Ib''). The signature for a $W\rightarrow \ell+\nu$ event is a lepton with high transverse momentum and large missing transverse momentum in the event, due to the undetected neutrino. In the electron channel, we select candidate events with the primary lepton in the CEM. In the muon channel, the lepton candidate is required to have stubs in the CMU, CMP or CMX. These conditions specify what is referred to here as the ``central lepton'' sample. Two samples of $Z\rightarrow e^++e^-$ and $Z\rightarrow \mu^++\mu^-$ are also used for tuning the simulation. The details of the trigger requirements can be found in \cite{wmassrun1}. The integrated luminosity is 80$\pm$4 pb$^{-1}$. 

The missing transverse momentum is inferred from the energy imbalance in the event. For this purpose, a recoil-energy vector $\vec{u}$ is defined as the vector sum of the transverse energies of all calorimeter towers (including both electromagnetic and hadronic, up to $|\eta|<3.6$), except the ones identified as part of the electromagnetic clusters associated with the primary leptons:
\begin{equation}
{\boldmath \vec{u}}= \sum_{i~{\rm not}~\ell} {E}_i \sin\theta _i {\hat{n}}_i,
\label{equ:recoildefinition}
\end{equation}
where $\hat{n}_i$ is a transverse unit vector pointing to the center of each tower and $\sin\theta_i$ is computed using the $z$-vertex closest to the electron track, or using the electron track $z_0$ if there is no $z$-vertex within 5 cm of the electron track. The vector $\vec{u}$ is a measure of the calorimeter's response to jets and particles recoiling against the $W$. Thus, the missing transverse energy (identified with the transverse momentum of the neutrino) is derived as $\not \!\!\vec{E}_T = -(\vec{P}_T^{\ell} + \vec{u})$, where $\vec{P}_T^{\ell}$ denotes the muon transverse momentum ($p_T$) or the electron transverse energy ($E_T$). The modulus ($u$) of the recoil vector is an estimator of the $W$ boson transverse momentum and it is used to select different ranges of the $W$ boost. 

The analysis uses the transverse mass ($M_T$), which is analogous to the invariant mass except only the transverse components of the four-momenta are used. $M_T$ is determined from the data as
\begin{equation}
M_T=\sqrt{2 P_T^\ell \not\!\!{E}_T (1-cos\Delta\phi^{\ell\nu})},
\label{equ:tmass}
\end{equation}
where $\Delta\phi^{\ell\nu}$ is the angle in the $r-\phi$ plane between the transverse momentum of the lepton and the missing energy.
 
Several selection criteria are chosen to isolate a sample of well measured electrons and muons and reduce the backgrounds. For candidates in the $W\rightarrow e+\nu$ sample, we select electrons with $E_T>$ 25 GeV and with the $p_T$ of the associated track greater than 15 GeV/$c$. Events are accepted only if $\not\!\!\!{E}_T>$ 25 GeV. We require a well measured track (crossing all eight super-layers of the CTC and with more than 12 stereo hits attached). To exploit the projective geometry of the CDF detector, the event vertex reconstructed with the VTX is selected to be within 60 cm in $z$ from the origin of the detector coordinates. Fiducial requirements are applied to ensure that candidates are selected in regions of well understood efficiency and performance of the detector. 
To remove $Z$-boson events from the $W$ sample a search is made for a partner electron in the central (CEM), plug (PEM), or forward (FEM) calorimeter. Partner electrons are sought with cluster transverse energies greater than 20 GeV, 15 GeV and 10 GeV in the CEM, PEM and FEM respectively. Tracks with transverse momentum greater than 10 GeV/$c$ and opposite sign to the primary electron are also considered. The event is rejected if the invariant mass of the primary electron with the partner electron exceeds 60 GeV/$c^2$. The event is also rejected if the partner electron is pointing to any non-fiducial volume of the calorimeter,
as this may cause the cluster's energy to be mis-measured and consequently cause the invariant mass rejection to fail. 

In order to improve electron identification, 
 additional variables are used. They are the ratio of the hadronic to electromagnetic deposited energies ($E_{had}/E_{em}< 0.1$), the match between the extrapolated track and the measured position at the CES ($\Delta z_{CES}<5$ cm), the transverse CES shower shape \cite{cesshape} ($\chi^2_z<10$), and the track isolation ($ISO_{0.25} < 1$ GeV/$c$). The track isolation variable $ISO_{R}$ is defined as the total transverse momentum from tracks (unconstrained by the vertex position) of $p_T>$ 1 GeV/$c$, that lie within a cone of semi opening $R = \sqrt{(\Delta\eta)^2+(\Delta \phi)^2}$ centered on the lepton track and within 5 cm of the lepton $z$ vertex. 

For candidates in the $W\rightarrow \mu+\nu$ sample, the muon $p_T$ and the $\not\!\!\!{E}_T$ in the event are required to be greater than 25 GeV. The quality requirements on the tracks are the same as for the electrons. In addition, there are requirements on the impact parameter of the track ($|d_0|<0.2$ cm) and on the opening angle ($>$ 10$^\circ$) with the second high-$p_T$ track to remove cosmic rays. The muon identification is based on the presence of track-stubs in the muon systems and on the deposited energy of the candidates in the calorimeters. The deposited energy is required to be less than 2 GeV in the CEM and 6 GeV in the CHA. Furthermore, we require that the CTC track, extrapolated at the center of the muon chambers, and the track-stub reconstructed in the muon systems match to within 2 cm in the CMU or 5 cm in the CMP and CMX. The track isolation cut has not been applied to muon candidates since the muon sample is smaller in size and we have preferred a looser selection. The $Z$ removal rejects events where there is a second highest-$p_T$ ($>10$ GeV/$c$) track in the CTC, of opposite sign to the $\mu$ candidate and back-to-back in space (within 10$^\circ$), that has an invariant mass with the $\mu$ candidate greater than 50 GeV/$c^2$. 

The $Z$ samples are selected with the same $W$ selection criteria, except the $\not\!\!{E}_T$ is replaced with a partner high-$p_T$ lepton, and the $Z$ removal requirements are not applied. Moreover, the sample of $Z\rightarrow e^++e^-$ used for the tuning of the simulation has two CEM electrons, both passing electron ID cuts. This choice removes almost all of the QCD background.

A summary of the selection requirements and the number of surviving events is shown in Tables~\ref{tab:weleccuts} (electrons) and \ref{tab:wmuoncuts} (muons). The accepted samples consists of 22,235 $W\rightarrow\mu+\nu$ candidates and 41,730 $W\rightarrow e+\nu$ candidates, divided in four recoil ranges.
\begin{table}[h]
\begin{center}
\begin{tabular}{|lr|}
\hline
Criterion & $W$ events after requirements \\
\hline
Initial sample & 105,073 \\
Fiducial requirements & 75,135 \\
Good electron track  & 68,337 \\
$E_T^e> 25$ GeV  & 64,254 \\
$E_T^\nu > 25$ GeV & 54,409 \\
$u<100$ GeV	& 54,300 \\
$p_T^e>$15 GeV/$c$ & 52,573 \\
$M_T=50-100$ GeV/$c^2$ & 51,077 \\
Electron ID & 42,882 \\
$Z$ removal & 41,730 \\
\hline
$u<10$ GeV  & 31,363 \\
$10<u<20$ GeV & 7,739 \\
$20<u<35$ GeV & 2,033 \\
$35<u<100$ GeV & 595   \\
\hline
\end{tabular}
\caption{{\small Set of requirements applied to select the $W\rightarrow e+\nu$ data sample.}}
\label{tab:weleccuts}
\end{center}
\end{table}
\begin{table}[h]
\begin{center}
\begin{tabular}{|lr|}
\hline
Criterion & $W$ events after requirements \\
\hline
Initial ample & 60,607 \\
$E_T^{CEM}<2$ GeV and $E_T^{CHA}<6$ GeV & 56,489 \\
Not a cosmic candidate & 42,296 \\
Impact parameter $d_0<0.2$ cm & 37,310 \\
Track-muon stub match & 36,596 \\
Good muon track & 33,887 \\
$p_T>25$ GeV/$c$ & 29,146 \\
$E_T^\nu>$ 25 GeV & 25,575 \\
$u<70$ GeV & 25,493 \\
$Z$ removal & 22,877 \\
$M_T$=50$-$100 GeV/$c^2$ & 22,235 \\
\hline
$u<7.5$ GeV  & 13,813 \\
$7.5<u<15$ GeV & 5,910 \\
$15<u<30$ GeV  & 2,088 \\
$30<u<70$ GeV & 424 \\
\hline
\end{tabular}
\caption{\small{Set of requirements applied to select the $W\rightarrow \mu+\nu$ data sample.}}
\label{tab:wmuoncuts}
\end{center}
\end{table}
\\
\begin{center}{\bf IV. MEASUREMENT METHOD}\end{center}
Ideally one would like to fit the distribution of $\cos\theta_{CS}$ for the coefficients $\alpha_1$ and $\alpha_2$ of Eq.~(\ref{equ:qcddist}). However, since the neutrino coming from the $W$ decay is undetected, the kinematics of the decay are not completely reconstructed and it is not possible to perform a boost into the $W$ rest frame and uniquely determine $\cos\theta_{CS}$. The finite width of the $W$ boson makes it difficult to solve the equations of the $W$ two-body decay. Even if the mass of the $W$ were known on an event by event basis and the detector had perfect resolution, the unknown longitudinal component of the neutrino momentum would leave a sign ambiguity in determining $\cos\theta_{CS}$.

\begin{figure}[h]
\begin{center}
\epsfig{file=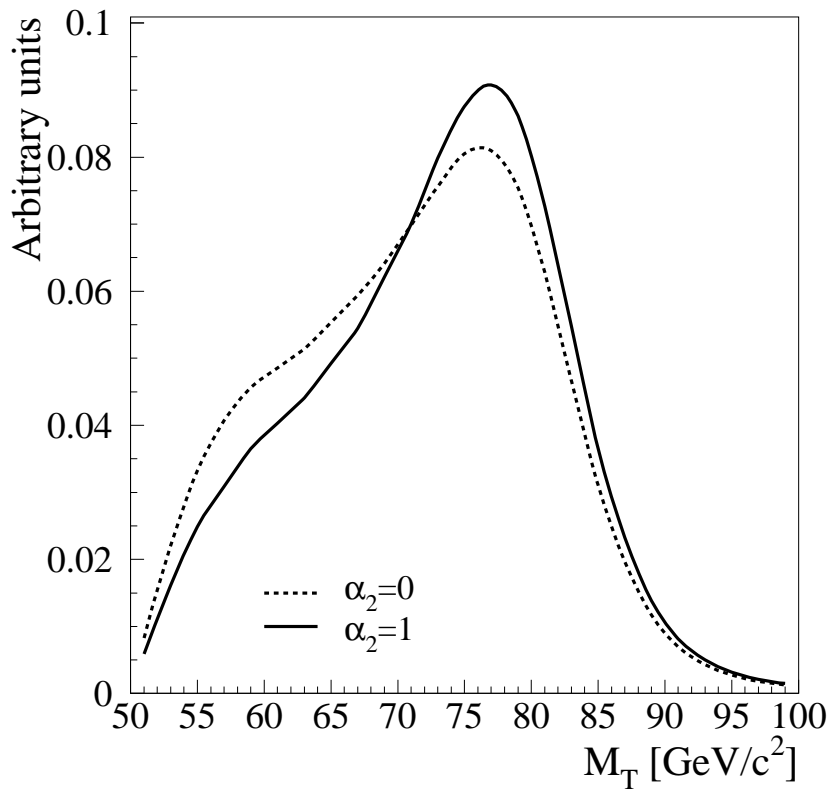,width=12cm,angle=0}
\caption{{\small Example of the sensitivity of the $M_T$ distribution to $\alpha_2$. Here $\alpha_2$ has been set to 0 and 1, and $p_T^W$ is less than 20 GeV/$c$.}}
\label{fig:correlation}
\end{center}
\end{figure}

This measurement therefore exploits the relationship between the transverse mass of the $W$ and the lepton polar angle on a statistical basis, {\it i.e.} by using the shape of the $M_T$ distribution. A similar technique has been successfully applied in \cite{d0paper} to measure $\alpha_2$ from $W\rightarrow~e+\nu$ decays. 
Fig.~\ref{fig:correlation} shows an example of how the distribution of the transverse mass of the $W$ changes with different values of $\alpha_2$. Also, since $M_T$ does not contain any information on the longitudinal boost of the $W$ boson, it is affected by $\alpha_1$ (the forward-backward lepton decay asymmetry term) only through residual effects of the geometrical acceptance of the detector.

The parameter $\alpha_2$ is determined by fitting the $M_T$ distribution to a set of Monte Carlo generated templates, each with a different value of $\alpha_2$. A binned log-likelihood method is applied to find the best estimate for $\alpha_2$. The procedure is repeated selecting different regions of the transverse momentum of the $W$ boson.
\\
\begin{center}{\bf V. {\boldmath MONTE CARLO SIMULATION OF $W$ PRODUCTION AND DECAY}}\end{center} 
A fast Monte Carlo (MC) generator and a parametrization of the detector response have been used in this analysis to simulate $W$ events at CDF \cite{wmassrun1}. The event generator is based on a leading order calculation of $W$ production and leptonic decay in quark-antiquark annihilation, including final state QED radiation \cite{BKgener,photos1,photos2,photos3}. The distribution of momenta of the quarks is based on the MRS-R2 \cite{mrs} set of parton distribution functions (PDFs). The generated $W$ boson is then Lorentz-boosted, in the center-of-mass frame of the quark-antiquark pair, to a specific transverse momentum $p_T^W$. This measurement uses a broad range of $p_T^W$, including events at low $p_T^W$, where theoretical calculations are not reliable. The spectrum of $p_T^W$ as a function of the $W$ boson rapidity is therefore derived from $p_T^Z$ (the $p_T$ of a $Z$ boson $-$ determined experimentally from $Z\rightarrow e^++e^-$, $\mu^++\mu^-$ events) after correcting it by the theoretical prediction for $p_T^W/p_T^Z$. There is no physics simulation of the recoiling jets, instead we model directly the detector response to the recoil against a $W$ boson. The parametrization of the detector response and the modeling of the $W$ boson recoil up to 20 GeV/$c$ is described in detail in \cite{wmassrun1}. We have tuned the parameters of the model to describe the range of $p_T^W$ up to 100 GeV/$c$. Overall, the MC tuning performed for this analysis involves:\\
$-$ the effects of QCD on the lepton angular distribution,\\
$-$ the parametrization of the $Z$ transverse momentum spectrum, up to\\ $\qquad p_T^Z$ = 100 GeV/$c$, and\\
$-$ the detector response to the recoil against high-$p_T$ $Z$ and $W$ bosons.\\

\begin{center}{\bf A: Effects of QCD on the lepton angular distribution}\end{center}
The QCD effects on the lepton angular distribution are implemented with an event weighting procedure in the simulation. Leptons from $W$ decays, generated with a tree-level quark-antiquark annihilation, have a purely $V-A$ angular spectrum with a very small distortion due to the final state photon emission. Therefore, events are first unweighted by $1/(1-Q \cos\theta)^2$, where $\theta$ is the lepton polar angle in the parton frame and $Q$ is the lepton charge. This effectively factors out any small distortion of the spectrum with respect to a parabola. Events are then assigned the appropriate weight ($w_{QCD}$), where $w_{QCD}$ is defined as a function of the lepton angles ($\theta_{CS}$, $\phi_{CS}$) in the Collins-Soper $W$ boson rest frame:
\begin{eqnarray}
w_{QCD}(\theta_{CS},\phi_{CS}) & = & 1+\cos^2\theta_{CS}+\frac{1}{2}A_0(1-3\cos^2\theta_{CS})+ \nonumber \\
& &+\frac{1}{2}A_2 \sin^2\theta_{CS}\cos2\phi_{CS}+ \nonumber \\
& & +A_3 \sin\theta_{CS} \cos\phi_{CS}-Q A_4\cos\theta_{CS}.  
\label{equ:wnlo}
\end{eqnarray}

Eq.~(\ref{equ:wnlo}) describes the angular modulation induced by the effects of QCD as expressed also in Eq.~(\ref{equ:qcdangular}), except for the terms with $A_{1,5,6,7}$; here they are set to zero, corresponding to the Standard Model prediction in the accessible $p_T^W$ range. The coefficients $A_2$ and $A_3$ are kept in the angular distribution and assigned the SM dependence with $p_T^W$, calculated in \cite{mirkes}. Notice that the angular coefficients to $A_2$ and $A_3$ cancel out when integrating analytically over $\phi_{CS}$ between 0 and $2\pi$. Nevertheless, detector acceptance effects introduce a small residual dependence in the polar-angle spectrum.

\begin{figure}[h]
\begin{center}
\epsfig{file=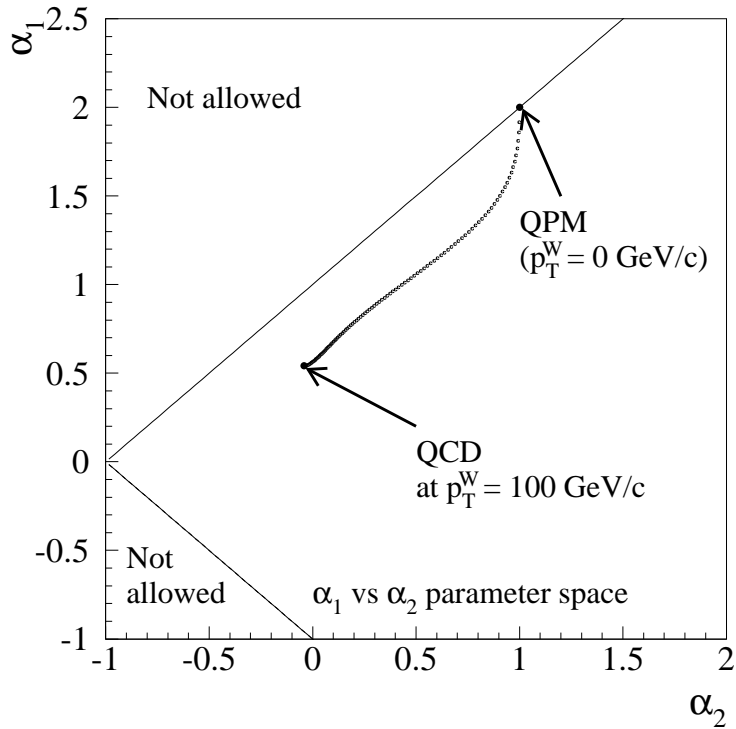,width=11cm,angle=0}\\   
\caption{\small{The $\alpha_1$ vs. $\alpha_2$ parameter space. The regions marked with ``not allowed'' are where the combination of $\alpha_2$ and $\alpha_1$ gives unphysical negative weights to the differential cross section. The dotted line shows the values of $\alpha_1$ and $\alpha_2$ at different $p_T^W$ between 0 and 100 GeV/$c$.}}
\label{fig:theo_weight}
\end{center}
\end{figure}

In Eq.~(\ref{equ:wnlo}), $w_{QCD}$ can take negative values if $A_0$ and $A_4$ (or, equivalently, $\alpha_2$ and $\alpha_1$) are varied independently in the procedure of fitting for the best parameters. Fig.~\ref{fig:theo_weight} shows the allowed parameter space for $\alpha_2$ and $\alpha_1$. The diagonals in the plot correspond to the requirement:
\begin{equation}
(1+\alpha_2 \cos^2\theta_{CS}\pm\alpha_1 \cos\theta_{CS})\geq 0,
\label{equ:weight_boundaries}
\end{equation}
for $\cos\theta_{CS} =\pm 1$. The point ($\alpha_1, \alpha_2)=(2, 1)$ is the Quark Parton Model (QPM) limit in the case that the sea quark contribution is neglected, and it has a vanishing cross section at $\theta_{CS}=\pm 180^\circ$, as described by the $V-A$ lepton angular distribution. The dotted line is the relationship between $\alpha_2$ and $\alpha_1$ (at different $p_T^W$ up to 100 GeV/$c$), expected from the SM including QCD corrections. To prevent $w_{QCD}$ from taking negative values, $\alpha_1$ and $\alpha_2$ are varied only within the allowed region. Note that the sea quark contribution to $\alpha_1$ is correctly taken into account in the Monte Carlo.

Because this is an event-weighting procedure, it does not correspond to the inclusion of QCD corrections to the generated events: the large-$p_T^W$ $W$ events still have to be introduced by hand, by imposing a transverse momentum distribution. 
\\
\begin{center}{\bf \boldmath B: $Z$ transverse momentum spectrum}\end{center}
Prior to the determination of the $Z$ transverse momentum distribution, the Monte Carlo simulation is tuned and checked against the $Z\rightarrow e^++e^-$ and $Z\rightarrow\mu^++\mu^-$ invariant mass distributions from the data. In the electron channel, the Monte Carlo simulation reproduces the data with an input $Z$-mass %($M_Z^{CDF}$) 
equal to the world-average \cite{pdg} within a scale factor of 1.0002$\pm$0.0009. In the parametrization of the energy resolution of the CEM:
\begin{equation}
\frac{\sigma_E}{E}=\frac{13.5\%}{\sqrt{E_T}}\oplus\kappa,
\label{equ:cemresolution}
\end{equation}  
we use $\kappa = (1.23\pm0.26)$\%. The $\kappa$ term accounts for residual gain variations not corrected by the calibration procedure and is obtained from a fit to the $Z$ invariant mass peak.

There is a small non-linearity correction to extrapolate the energy-scale corrections from electrons at the $Z$ pole to the energies typical of a $W$ decay. The average $E_T$ for  electrons coming from $Z$ decay is about 4.5 GeV higher than the $E_T$ for $W$ decay. The non-linearity over a small range of energies can be expressed with a slope as
\begin{equation}
S_E(W) = S_E(Z)\cdot \left[ 1+\xi\Delta E_T\right],
\end{equation}
where $S_E(Z)$ is the measured scale at the $Z$ pole, $\xi$ is the non-linearity factor, and $\Delta E_T$ is the difference in the average $E_T$ between $Z$ and $W$ electrons. The estimate of $\xi$ is derived by looking at $E/p$ distributions from the $W$ data and comparing them to the Monte Carlo simulation in separate regions of $E_T$.  We estimate $\xi$ to be
\begin{equation}
\xi =-0.00027\pm0.00005 ({\rm stat})~{\rm GeV}^{-1}.
\end{equation}

For muons, we use a momentum resolution of
\begin{equation}
\sigma(1/p_T) = (0.097\pm0.005)\times 10^{-2}~ ({\rm GeV}/c)^{-1},
\end{equation}
and the reconstructed $Z$ mass peaks in the data and MC match with a ratio of central values of 1.0008$\pm$0.0011. With these inputs, the Monte Carlo simulation reproduces correctly the peak position and width of the invariant mass distribution of electron and muon pairs from $Z$ bosons. 

Since the QCD corrections to $Z$ production are not included in the Monte Carlo simulation, the transverse momentum of the $Z$ bosons needs to be determined from data. The $p^Z_T$ distribution is generated in the Monte Carlo simulation using the following $ad$ $hoc$ four-parameter functional form:
\begin{equation}
f(p_T^Z)=\frac{x^{P_4}}{\Gamma (P_4+1)}\left[(1-P_1)P_2^{P_4+1} e^{-P_2x}+P_1P_3^{P_4+1}e^{-P_3 x}\right] ; \quad x=p_T^Z/(50.0~{\rm GeV}/c).
\label{equ:Zptfunctional}
\end{equation}

\begin{figure}[h]
\begin{center}
\begin{tabular}{ll}
\epsfig{file=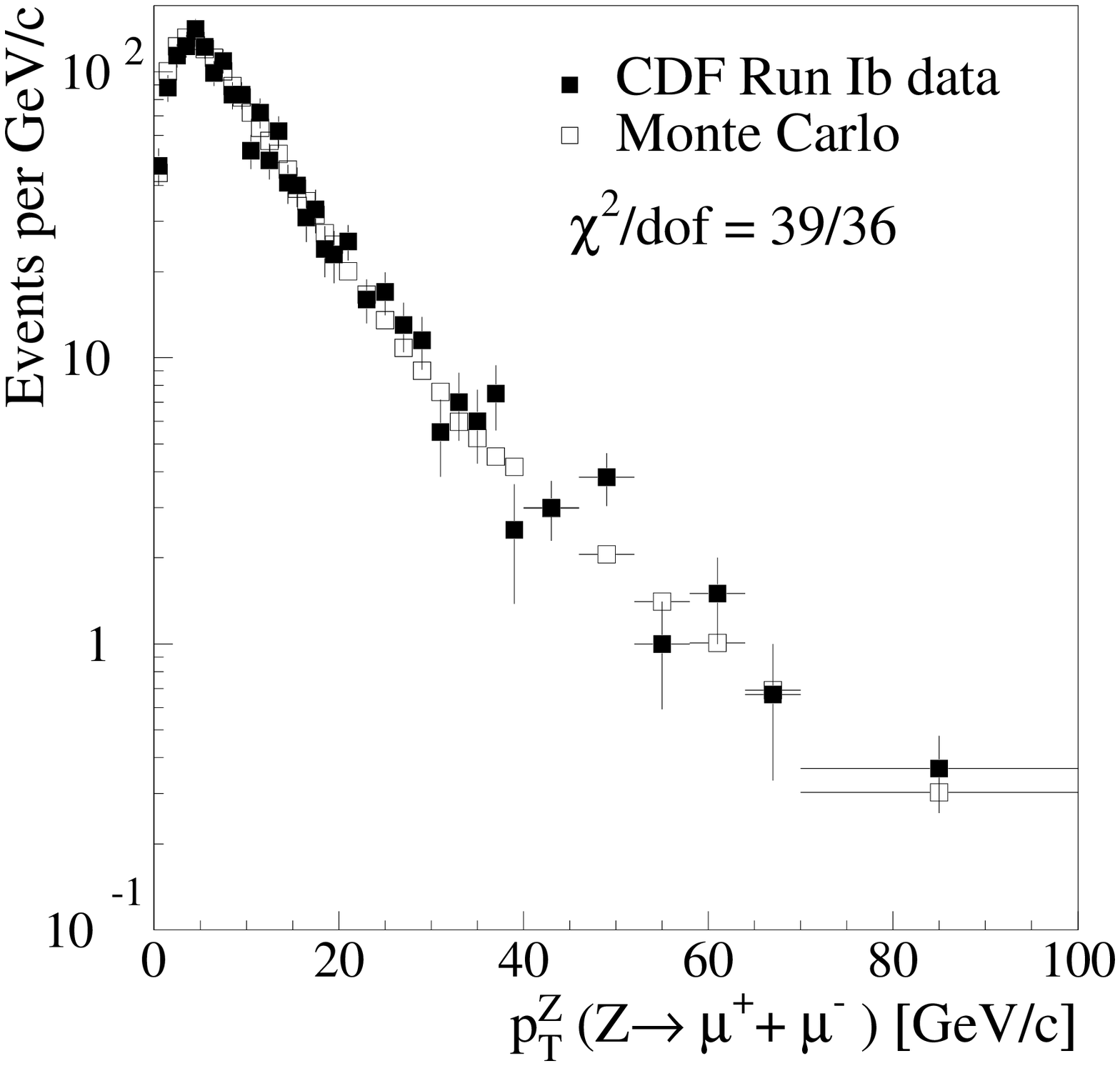,width=7.8cm,angle=0} &
\epsfig{file=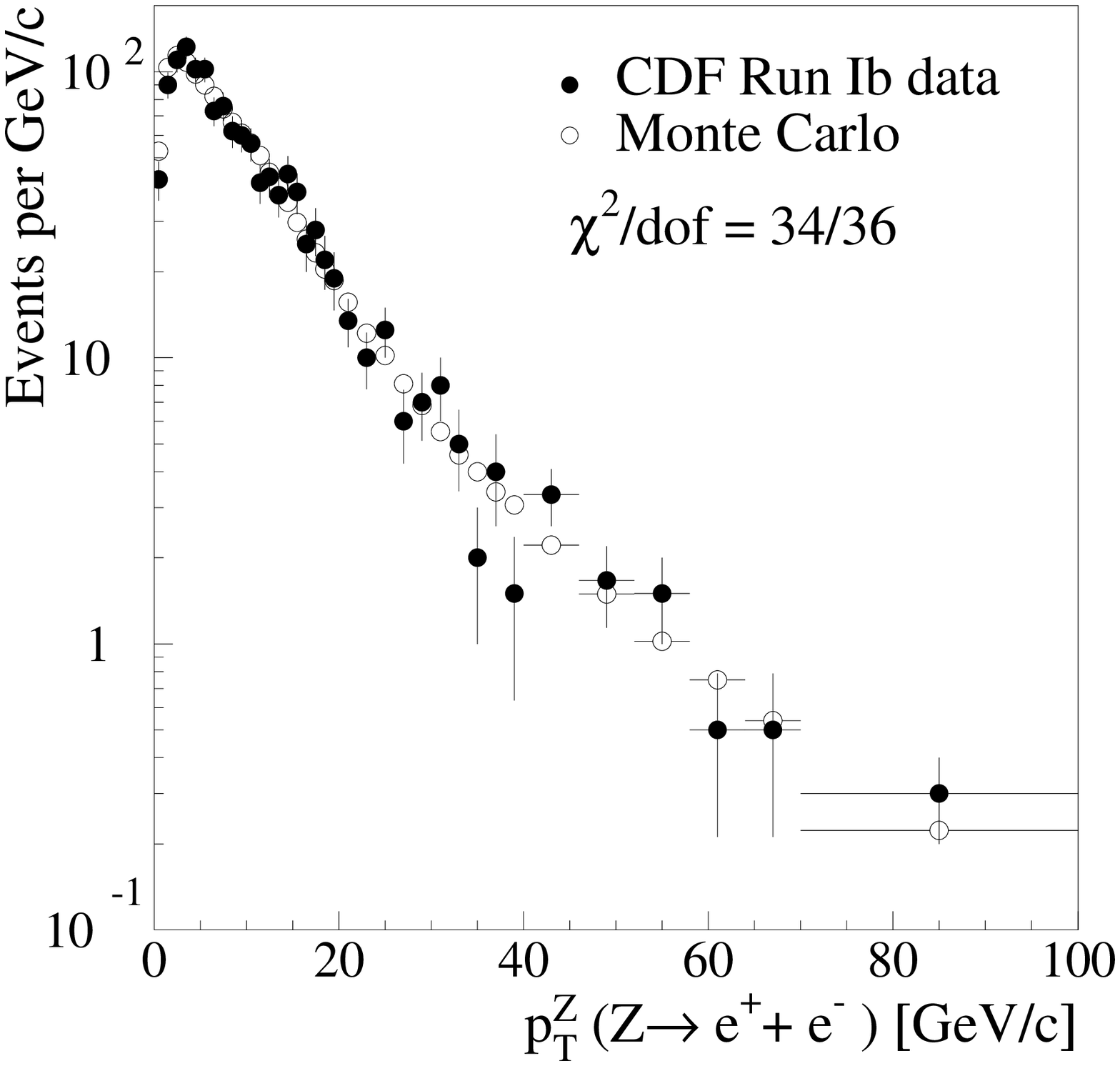,width=7.8cm,angle=0} \\
\hspace{3.2cm}\Large(a)\normalsize  &\hspace{3.2cm}\Large(b)\normalsize \\
\end{tabular}
\end{center}
\caption{\small{Distributions of $p_T^Z$ from $Z\rightarrow\mu^++\mu^-$ data (a) and $Z\rightarrow e^++e^-$ data (b) compared with the simulation.}}
\label{fig:ptz_frommm_newfit}
\end{figure}

The parameters  $P_{1,..,4}$ are determined from a fit to the observed $p_T^Z$ distribution and then corrected to account for the difference between the {\it observed} and the {\it generated} $p_T^Z$ spectrum. Since the difference between the two spectra is very small, the unfolding of the effect of the reconstruction is obtained by considering the ratio between them, as predicted by the detector simulation. We determine the $p_T^Z$ distribution using  separately $Z\rightarrow \mu^++\mu^-$ and $Z\rightarrow e^++e^-$ data, 
 and the average is used as the $p_T^Z$ spectrum that is input to the Monte Carlo simulation. The uncertainty on the average is used to evaluate the systematic uncertainty due to the transverse momentum spectrum determination. Fig.~\ref{fig:ptz_frommm_newfit} shows the $p_T^Z$ distribution of $Z\rightarrow \mu^++\mu^-$ and $Z\rightarrow e^++e^-$ data. The $p_T^Z$ spectra are compared with the simulation where the parameters have been fit to the data. There is a good agreement between data and Monte Carlo simulation and the $\chi^2$ values, normalized per degree of freedom, are very close to 1.
\\
\begin{center}{\boldmath \bf C: Detector response to the recoil against high-$p_T$ $Z$ and $W$ bosons}\end{center}
An estimate of the $W$ boost in the transverse plane comes from the measurement of the calorimeter response to jets and particles recoiling against the $W$. The definition of the recoil-energy vector $\vec{u}$ is given in Eq.~(\ref{equ:recoildefinition}). The modeling of $\vec{u}$ in terms of the $W$ boson transverse momentum is called the ``recoil model'' and it is implemented in the Monte Carlo simulation of the event. The recoil model is derived using the observed recoil against $Z$ bosons, whose kinematics are completely determined by the two leptons. The assumption is made that the recoil against $Z$ bosons can be extended to model $W$ events, since the $W$ and $Z$ bosons share a common production mechanism and are close in mass. We summarize below the key elements of the recoil model and show how the simulation describes the data after fitting the  model's parameters to the high-$p_T$ $Z$ boson data.
  
\begin{center}{\it 1. Recoil Model}\end{center}
The direction of $\vec{p}_T^{~Z}$ measured from the reconstructed decay leptons and the perpendicular to it form a base in the $r-\phi$ plane on which the recoil vector  $\vec{u}$ can be projected: $\vec{u}$= ($u_{||}$, $u_\perp$). The values of $u_{||}$ and $u_\perp$ are functions of $p_T^Z$ (addressed here as ``response functions'') with a certain smearing. The smearings are to a good approximation Gaussian distributions \cite{mythesis}, so that $u_{||}$ and $u_{\perp}$ can be parametrized as Gaussians with variable mean and width:
\begin{equation}
\left(   
\begin{tabular}{c}
$u_{||}$ \\
$u_{\perp}$ \\
\end{tabular}
\right)
=
\left(
\begin{tabular}{c}
$G[f_{||}(p_T^Z),\sigma _{||}(p_T^Z)]$ \\
$G[f_\perp(p_T^Z),\sigma _\perp(p_T^Z)]$ \\
\end{tabular}
\right).
\label{equ:umodel}
\end{equation}

\begin{figure}[tp]
\begin{center}
\begin{tabular}{rl}
\hspace{-0.5cm}\epsfig{file=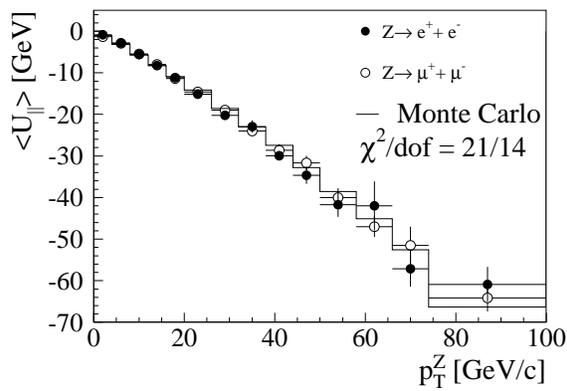,width=8.3cm,angle=0} &
\hspace{-0.8cm}\epsfig{file=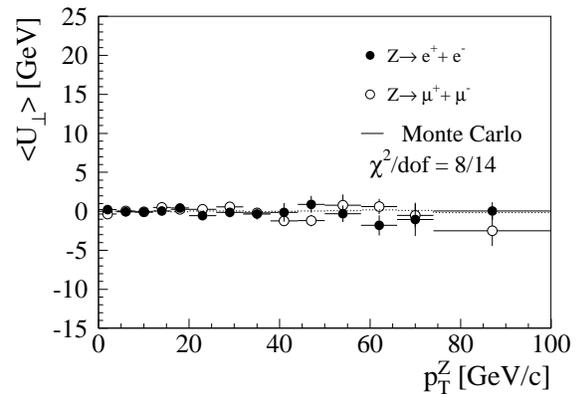,width=8.3cm,angle=0} \\   
&\hspace{-5.2cm}\Large(a)\normalsize  \hspace{7.2cm}\Large(b)\normalsize \\
\hspace{-0.5cm}\epsfig{file=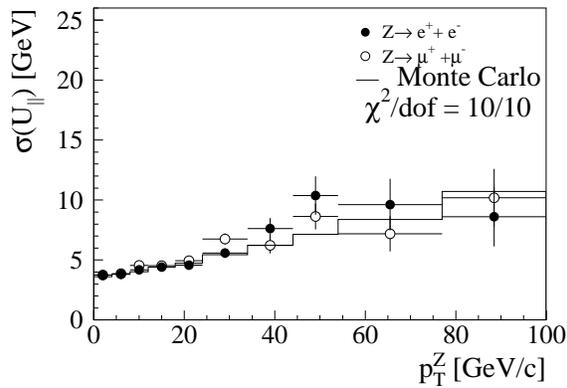,width=8.3cm,angle=0} &   
\hspace{-0.8cm}\epsfig{file=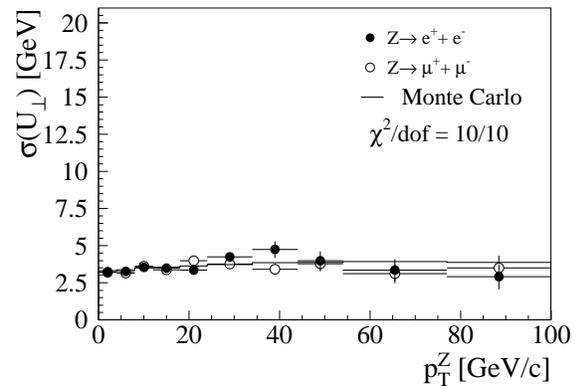,width=8.3cm,angle=0} \\   
&\hspace{-5.2cm}\Large(c)\normalsize  \hspace{7.2cm}\Large(d)\normalsize \\
\end{tabular}
\end{center}
\caption{\small{(a) and (b) Comparison of the data with the simulation for the recoil response components $u_{||}$ and $u_\perp$ versus $p_T^Z$. (c) and (d) the resolutions $\sigma(u_{||})$ and $\sigma(u_\perp)$ versus $p_T^Z$.}}
\label{fig:zrecoil_u}
\end{figure}

\begin{center}{\it 2. Response functions}\end{center}
The response function $f_{||}$ is well described by a second order polynomial in the $Z$ transverse momentum measured from the reconstruction of the decay leptons. The parameters for $f_{||}(p_T^Z)$ are obtained from a fit to $Z\rightarrow e^++e^-$ and $Z\rightarrow\mu^++\mu^-$ data and the function is corrected for a small difference between the true $p_T^Z$ and the observed $p_T^Z$ $-$ which is measured from the two leptons' momentum vectors $-$ to feed the correct parameters to the simulation. Fig.~\ref{fig:zrecoil_u}(a) shows the average of $u_{||}$, which is the response function for the parallel component, together with the simulation after fitting for the parameters of $f_{||}$. $u_{||}$ is on average smaller than $p_T^Z$, due to the gaps in the calorimeter and inefficiency in the reconstruction of the total energy deposited. Nonetheless, measuring $u_{||}$ provides an estimate of $p_T^Z$ (or ultimately $p_T^W$).

The response function $f_\perp(p_T^Z)$ is consistent with zero within the statistical uncertainty, as expected since $u_\perp$ is the recoil projection perpendicular to $p_T^Z$. The average of $u_\perp$ is shown in Fig.~\ref{fig:zrecoil_u}(b).

\begin{center}{\it 3. Resolutions}\end{center}
The resolution of the recoil vector components depends on the underlying event and the jet activity. $\sigma_{||}$ and $\sigma_{\perp}$ are parametrized in the form:
\begin{equation}
\left(
\begin{tabular}{c}
$\sigma_{||}$ \\
$\sigma_{\perp}$ \\
\end{tabular}
\right)=
\begin{tabular}{c}
$\sigma_{mbs}(\sum E_T)\quad\times$ \\
\end{tabular}
\left(
\begin{tabular}{c}
$P_{2,||}(p_T^Z)$\\
$P_{2,\perp}(p_T^Z)$\\
\end{tabular}
\right),
\label{equ:recmodel}
\end{equation}
where $P_{2,||}$ and $P_{2,\perp}$ are second order polynomials in $p_T^Z$, whereas $\sigma_{mbs}$ contains the underlying event contribution and is modeled by minimum bias events. In Eq.~(\ref{equ:recmodel}), $\sigma_{mbs}$ is expressed as a function of the total transverse energy $\sum E_T$, defined as the scalar sum of tower transverse energies:
\begin{equation}
\sum E_T= \sum_{i~{\rm not}~\ell^{\pm}} E_i \sin\theta _i.
\end{equation}

$\sum E_T$ is a measure of the total transverse energy in the event from all sources, excluding the primary lepton. The functional dependence of $\sigma_{mbs}$ versus $\sum E_T$ is calculated in \cite{wmassrun1}. The explicit $p_T^Z$ dependence in the polynomials is derived here from $Z$ data, using both electrons and muons. The parameters are then corrected for the dependence of the observed $p_T^Z$ versus the true $p_T^Z$, as done for the response functions. Fig.~\ref{fig:zrecoil_u}(c),(d) show the resolution of $u_{||}$ and $u_{\perp}$. The resolution $\sigma(u_{||})$ worsens at higher $p_T^Z$, due to increased jet activity in the event. The agreements between data and Monte Carlo simulation are good in all the plots and the $\chi^2$'s normalized per degree of freedom are close to 1.
\\
\begin{center}{\bf \boldmath D: $W$ transverse momentum distribution}\end{center}
To turn the $p_T^Z$ distribution into a $p_T^W$ distribution, the simulation applies two weighting functions. The first allows for the fact that the $p_T^Z$ distribution (as in Eq.~(\ref{equ:Zptfunctional})) is derived with a fit performed to data averaged over all rapidity values (with mean $|y|$=0.3). However, $W$ events need to be generated differentially in both $p_T$ and $y$. This weighting function is taken from a theoretical calculation of $\frac{{\rm d}^2\sigma}{{\rm d} p_T {\rm d}y}/\left<\frac{{\rm d}\sigma}{{\rm d}p_T}\right>_y$ \cite{wmassrun1}. 

The second weighting function turns the $p_T^Z$ distribution, generated with both $p_T$ and $y$ dependence, into a distribution for the transverse momentum of the $W$ boson. This is obtained from the theoretical calculation of $\left.\frac{{\rm d}^2\sigma}{{\rm d}y {\rm d}p_T}\right|_W/\left.\frac{{\rm d}^2\sigma}{{\rm d}y{\rm d}p_T}\right|_Z$ \cite{zwtheo, zwtheo2, zwtheo3, zwtheo4}. Resummed calculations are used for correcting the difference between the $W$ and the $Z$ $p_T$ distributions. The ratio is between 0.9 and 1.0 over the $p_T$ range of interest. Since this is a ratio, the uncertainty is expected to be small because of cancellation of systematics. Indeed, by varying the PDF, $\alpha_s$, or the type of calculation, the resulting uncertainty in $p^W_T$ is small in comparison to the uncertainty arising from the statistics of the $Z$ sample used to define the distribution \cite{zwrat1,zwrat2,zwrat3,zwrat4}. 

\begin{figure}[htp]
\begin{center}
\begin{tabular}{ll}
\epsfig{file=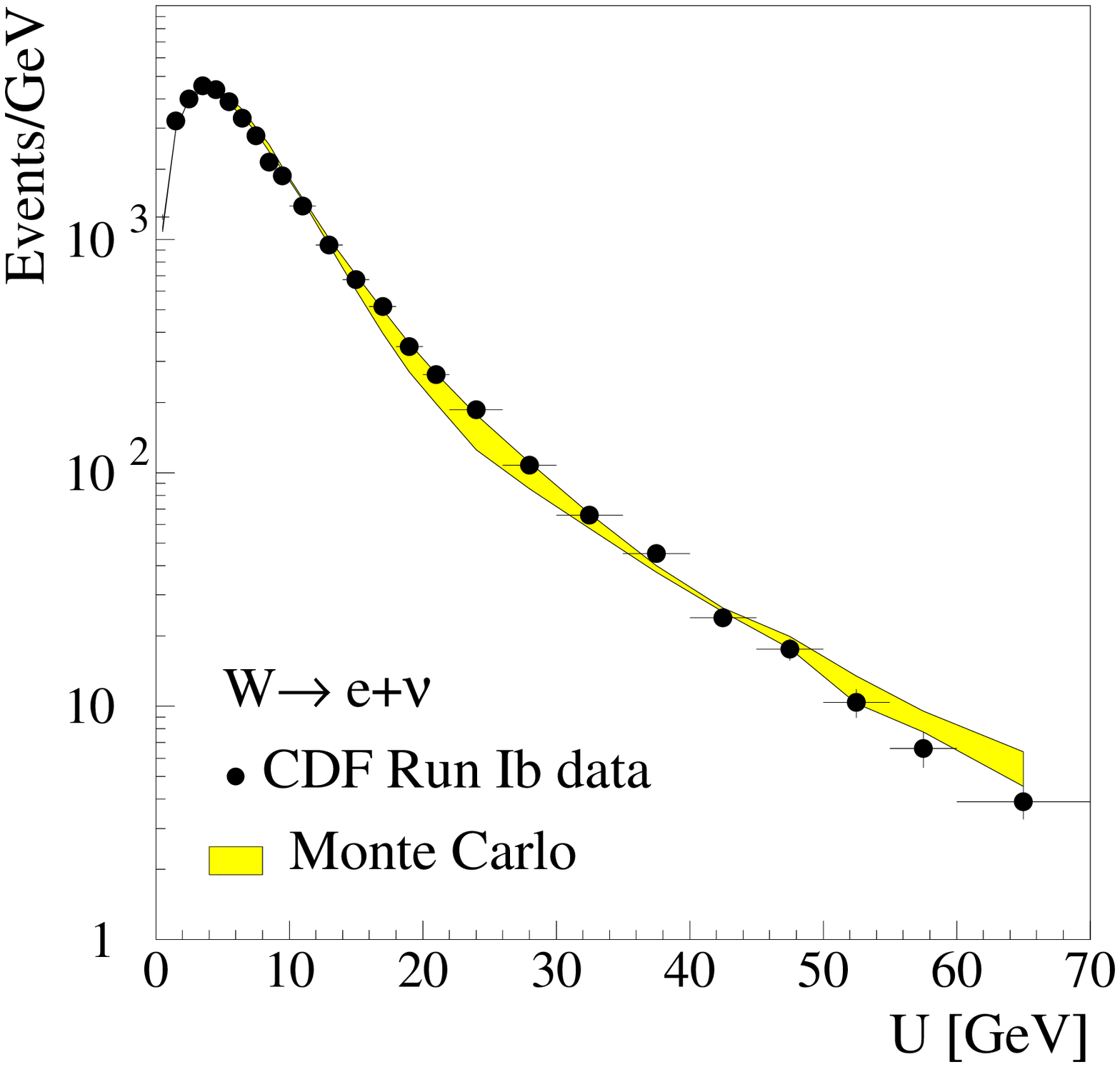,width=7.8cm,angle=0} &
\epsfig{file=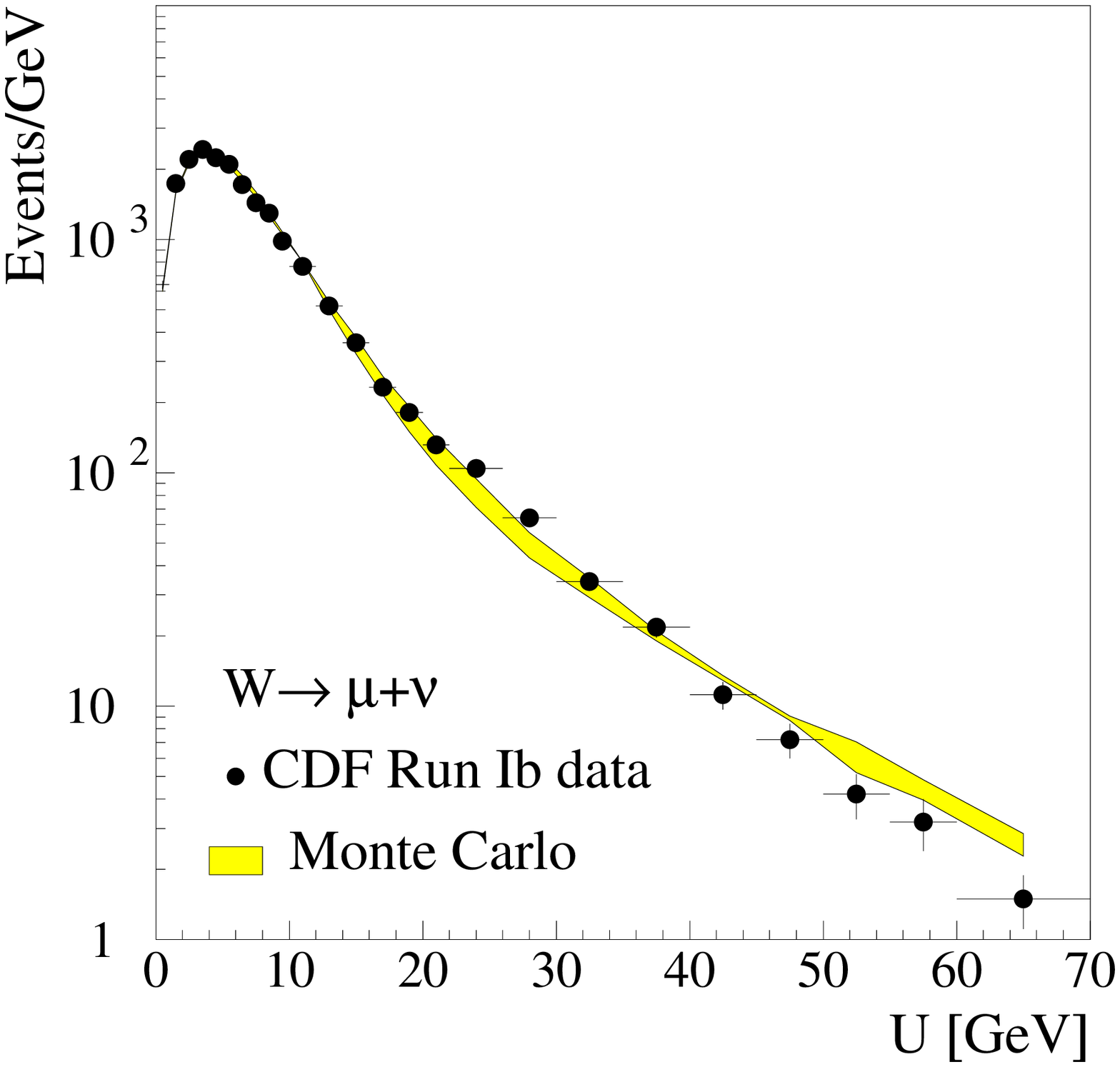,width=7.8cm,angle=0} \\ 
\hspace{3.2cm}\Large(a)\normalsize  &\hspace{3.2cm}\Large(b)\normalsize \\
\end{tabular}
\end{center}
\caption{\small{Distribution of the recoil against the $W$ boson compared with the simulation in $W\rightarrow e +\nu$ data (a) and  $W\rightarrow \mu+\nu$ data (b).}}
\label{fig:wrecoil_el}
\end{figure}

Although due to the undetected neutrino we cannot compare directly the $p_T^W$ spectrum in the simulation with the data, Fig.~\ref{fig:wrecoil_el} shows a comparison of the recoil against the $W$ in the electron and muon channel. The recoil includes the $p_T^W$ distribution as well as all the response and resolution parameters derived using the $Z$ sample. The shaded band corresponds to the uncertainty on the $p_T^Z$ spectrum only. Since the recoil model and the $p_T^Z$ spectrum are derived with a sample that is much smaller than the $W$ sample, there is a degree of freedom in optimizing the parameters to improve the agreement with $W$ data. However, we choose not to optimize the parameters using any of the $W$ boson distributions to prevent a possible source of bias when fitting the transverse mass distribution. We treat the statistical uncertainty of the recoil model and $p_T^Z$ spectrum as a source of systematic uncertainty for $\alpha_2$.

\begin{center}{\bf VI. BACKGROUNDS}\end{center}
There are three main sources of background to the $W\rightarrow \ell+\nu$ data sample of this analysis (where $\ell$ stands either for an electron or a muon):
\begin{itemize}
\item W$\rightarrow \tau+\nu$ events, with the $\tau$ subsequently decaying into a muon or electron and two neutrinos.
\item Z$\rightarrow \ell^++\ell^-$ events, where one of the leptons is not detected.
\item QCD dijet events, where a jet is wrongly identified as a lepton and the total energy in the event is incorrectly measured to give a $\not \!\!{E}_T$ signal. 
\end{itemize} 

There is a small background contribution from $t\bar{t}$ decays, which is estimated to be $\sim$25 events in the electron channel and $\sim$12 in the muon channel \cite{ttbarref} and affects the high recoil range only. The background from cosmic rays in the muon channel is approximately  0.2\% \cite{wmassrun1} of the total $W\rightarrow \mu+\nu$ candidates, with a flat $M_T$ distribution. This corresponds to a negligible contribution compared with the dominant backgrounds.

A shape for the transverse mass distribution is determined for each background source and added to the transverse mass distribution of the simulated $W$ events. For $t\bar{t}$ background the shape is taken from \cite{ttbarref2}.

\begin{center}{\bf \boldmath A: $W\rightarrow  \tau+\nu$ background}\end{center}
The background from $W\rightarrow\tau+\nu$ events, where the $\tau$ decays leptonically, is virtually indistinguishable from the $W\rightarrow e+\nu$ or $W\rightarrow\mu+\nu$ signal. The event generator used for the simulation of $W$ events in this analysis is capable of simulating $W\rightarrow\tau+\nu$, where the $\tau$ lepton is then decayed into $\mu+2\nu$ or $e+2\nu$. The background level is found to be approximately 2\% of the total $W$ sample, with softer charged lepton $p_T$ and $\not \!\!{E}_T$ spectra. The $W\rightarrow\tau+\nu$ background fractions are listed in Tables~\ref{tab:backelec} and \ref{tab:backmuon} for the electron and muon channel respectively. The shape of the transverse mass distribution is also taken from the Monte Carlo simulation of $W\rightarrow\tau+\nu$ events, separately for each of the $W$ boson recoil ranges.

\begin{center}{\bf \boldmath B: $Z\rightarrow\ell^++\ell^-$ background}\end{center}
$Z$ events enter the $W$ sample when one of the leptons is not detected (``lost leg'') and there is missing transverse energy in the event.

\begin{center}{\it 1. Electron channel}\end{center}
As part of the $W$ candidate selection procedure the primary electron is always required to have been detected in the central calorimeter. The $Z$ removal procedure ensures the rejection of events with a second oppositely charged high-$p_T$ track, or high-energy calorimeter cluster, and invariant mass of the electron-candidate pair compatible with a $Z$ boson decay ($M_{ee}> 60$ GeV/$c^2$). When the track associated with the second electromagnetic cluster is pointing to any non-fiducial volume of the calorimeter, the event is rejected irrespective of the invariant mass value. This ensures that the event would still be rejected if the second electron has emitted a photon and the invariant mass with the primary electron track falls outside the $Z$ invariant mass exclusion range. 

The Monte Carlo simulation is used to estimate the $Z$ background due to the inefficiency of the calorimeters in detecting the second leg, or when the second electron points beyond the coverage of the forward calorimeter ($|\eta|>$ 4.2). The total background level from $Z$ events in the electron channel is very small, and is listed in Table~\ref{tab:backelec}.
\begin{table}[h]
\begin{center}
\begin{tabular}{|l||cccc|r|}
\hline
       &           &             & Recoil [GeV] &             &     \\
 Type:  &     (0$-$10) & (10$-$20) & (20$-$35) & (35$-$100)  & All \\
\hline
$W\rightarrow\tau+\nu\qquad$ & 2.15  &1.74 & 1.31  & 1.57 & 2.01 \\
$Z\rightarrow e+(e)$         & 0.00  &0.02  & 0.12 & 0.39 & 0.01 \\
QCD jets                    & 0.23$\pm$0.11  &0.39$\pm$0.14  & 0.14$\pm$0.10 & 0.5$\pm$0.3  & 0.26$\pm$0.12  \\
$t\bar{t}$                  & 0.00           & 0.00          & 0.49$\pm$0.20 & 2.50$\pm$0.80  &0.06$\pm$0.02  \\
\hline
Total                       & 2.38$\pm$0.11  &2.15$\pm$0.14  & 2.06$\pm$0.22  &4.96$\pm$0.85 & 2.42$\pm$0.12 \\
\hline
\end{tabular}
\end{center}
\caption{\small{Summary of the backgrounds to $W\rightarrow e+\nu$ (as percentages of the $W$ candidate sample) in different $W$ recoil ranges. The uncertainty is negligible for $W\rightarrow \tau+\nu$ and $Z\rightarrow e+(e)$.}} 
\label{tab:backelec}
\end{table}

\begin{center}{\it 2. Muon channel}\end{center}
The event selection applied in this analysis removes events with opposite sign tracks (found in the CTC) that combine with the identified muon to give an invariant mass greater than 50 GeV/$c^2$. The number of $Z\rightarrow\mu^++\mu^-$ events not removed by the $Z$ selection criteria is consistent with zero when both muons pass through the fiducial tracking volume ($|\eta|<1$). 

However, a significant number of $Z$ events may enter the $W$ sample when one of the muons goes outside the fiducial tracking volume. About 20\% of $Z\rightarrow\mu^++\mu^-$ events have one of the muons outside $|\eta|<1$, either at the edge of the tracking volume ($|\eta|\sim 1.1$) or at higher $\eta$, beyond the coverage of the CTC. The estimate of the background in these cases is based on the simulation, which uses the tracking efficiency map determined using electrons detected in the calorimeter from the $W\rightarrow e+\nu$ sample. The background level found is of the order of 4\% and it is listed in Table~\ref{tab:backmuon}. The shape of the transverse mass distribution of lost-leg events is also derived from the Monte Carlo simulation.

\begin{table}[h]
\begin{center}
\begin{tabular}{|l||cccc|r|}
\hline
       &           &             & Recoil [GeV] &             &     \\
 Type: & (0$-$7.5)   & (7.5$-$15)    & (15$-$30)        & (30$-$70)    & All \\
\hline
$W\rightarrow\tau+\nu\qquad$  & 2.24           &1.94            &  1.63         & 2.37         & 2.11 \\
$Z\rightarrow \mu+(\mu)$      & 4.25           &4.00            & 3.67          & 2.95         & 4.11 \\
QCD jets                     & 0.45$\pm$0.19  & 0.79$\pm$0.29  & 0.81$\pm$0.52 & 1.40$\pm$1.18  & 0.59$\pm$0.26 \\
$t\bar{t}$                   & 0.00           & 0.00          & 0.19$\pm$0.09 & 1.89$\pm$0.70  &0.05$\pm$0.02  \\
\hline
Total                        & 6.94$\pm$0.19  &6.73$\pm$0.29 & 6.30$\pm$0.53  &8.61$\pm$1.37 & 6.86$\pm$0.26 \\
\hline
\end{tabular}
\end{center}
\caption{\small{Summary of the backgrounds to $W\rightarrow \mu+\nu$ (as percentages of the $W$ candidate sample) in different $W$ recoil ranges. The uncertainty is negligible for $W\rightarrow \tau+\nu$ and $Z\rightarrow \mu+(\mu)$.}} 
\label{tab:backmuon}
\end{table}

\begin{center}{\bf C: QCD background}\end{center}
Dijet events can pass the $W$ selection cuts if one of the jets is mis-identified as a lepton and one of them is incorrectly measured and gives a high missing-$E_T$ signal. This is referred to as QCD background. $W$ candidate events which are background from QCD would typically have the charged lepton or the neutrino predominantly back-to-back or collinear with the leading jet. Real $W$ events, on the other hand, have a nearly uniform distribution of the lepton-jet opening angle, at least for low $p_T^W$. For higher $p_T^W$, $W$ events also exhibit a slight tendency to have the leading jet, which is recoiling against the $W$, in the opposite direction to the charged lepton and the neutrino. 

\begin{figure}[h]
\begin{center}
\begin{tabular}{ll}
\epsfig{file=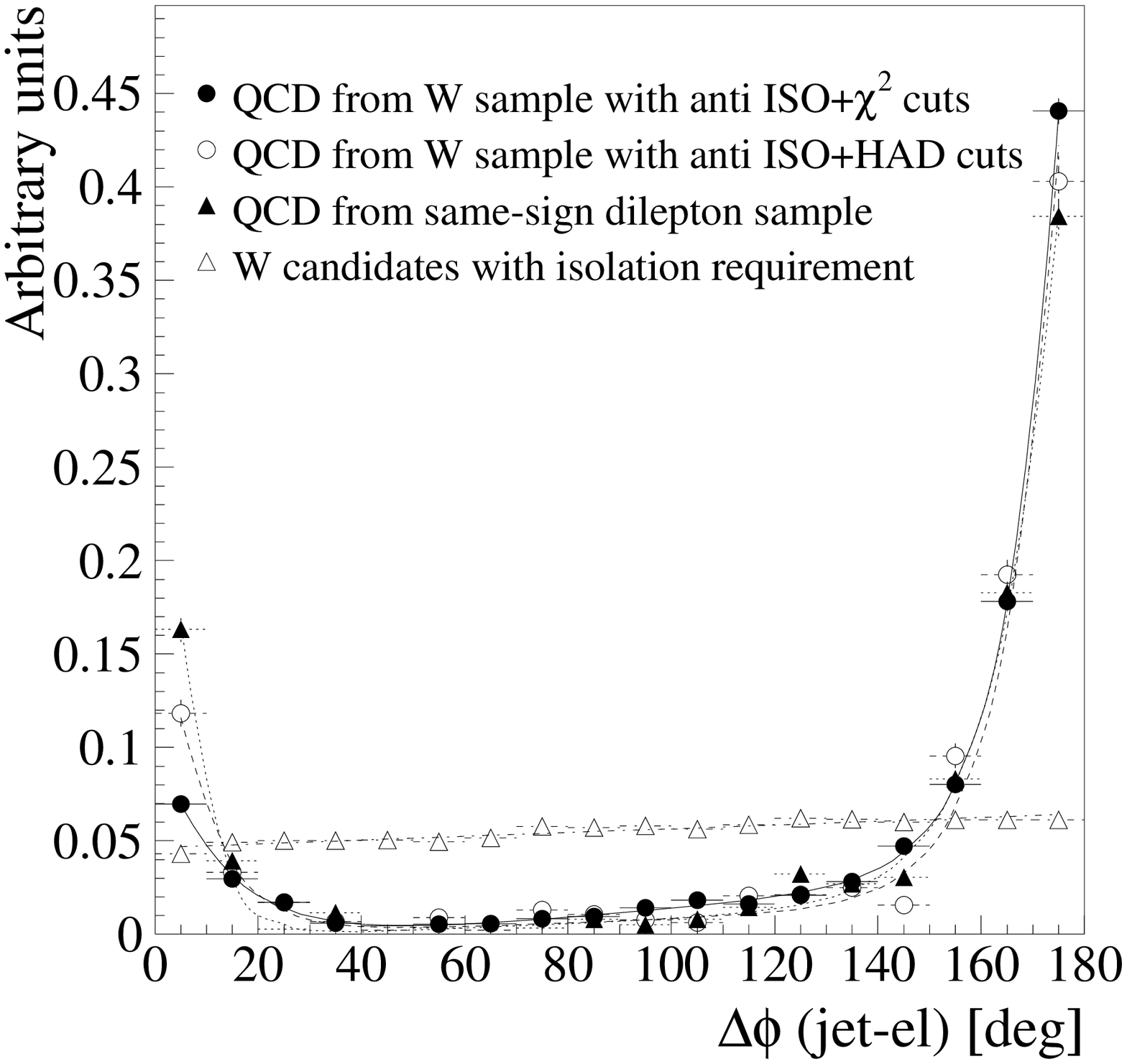,width=7.8cm,angle=0} &
\epsfig{file=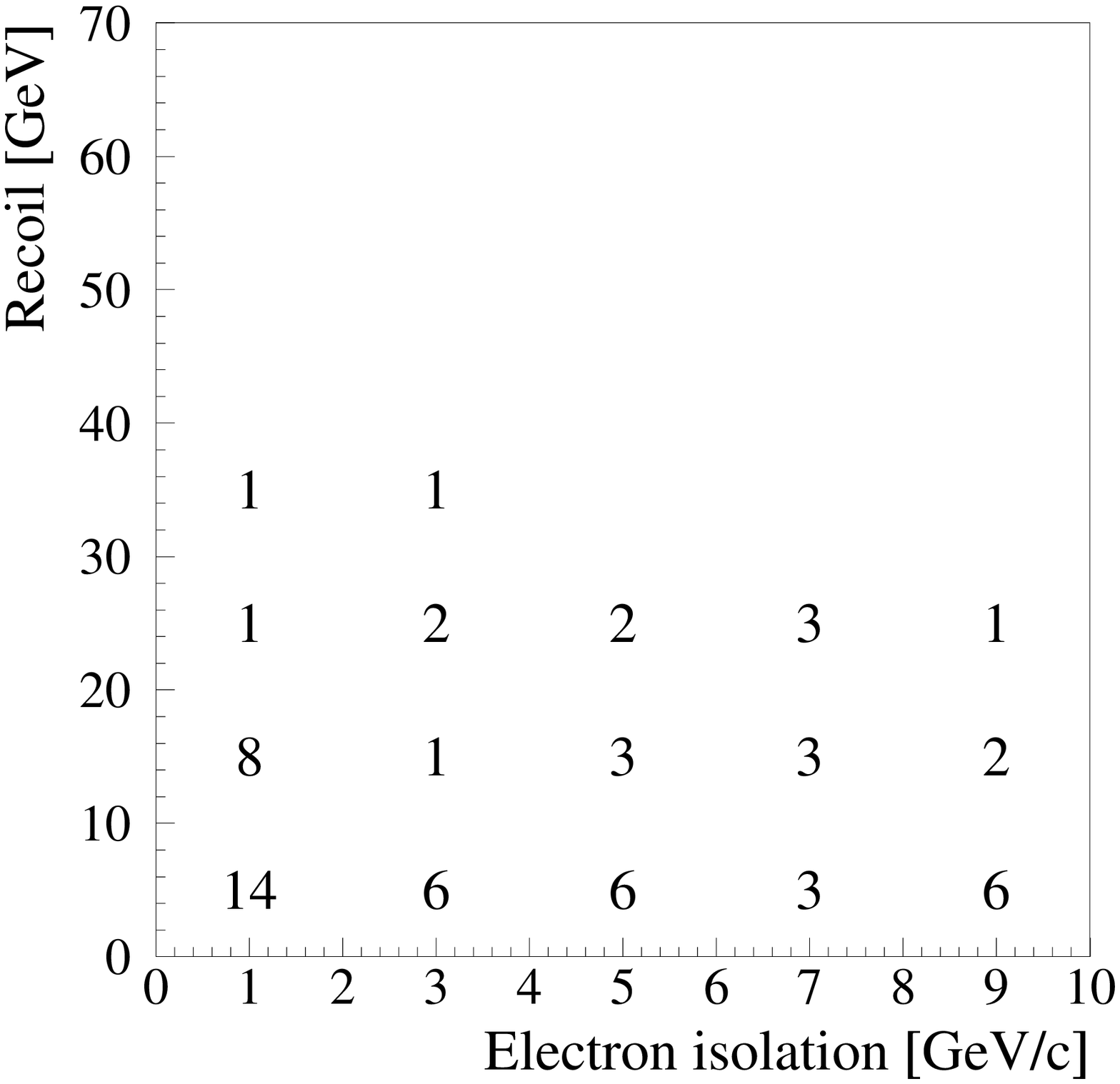,width=7.8cm,angle=0} \\
\hspace{3.2cm}\Large(a)\normalsize  &\hspace{3.2cm}\Large(b)\normalsize \\
\end{tabular}
\end{center}
\caption{\small{(a) Azimuthal angle between the electron candidate and the leading jet in the QCD samples and in the $W$-candidates sample. (b) Number of events in the plane of recoil versus isolation in the QCD-enriched sample, derived from the dilepton sample with a same-sign requirement and all the electron-ID cuts applied.}}
\label{fig:qcd_typical}
\end{figure}

\begin{center}{\it 1. Electron channel}\end{center}
Fig.~\ref{fig:qcd_typical}(a) shows the distribution of the opening angle in the $r-\phi$ plane between the electron and the leading jet. The leading jet is the highest energy jet in the event with energy of at least 5 GeV. The plot shows three samples enriched in QCD background together with the $W$ candidates sample. Two of the enriched QCD samples are derived by reversing the electron ID cuts in the $W$ preselection sample. The third is taken from dilepton events ($Z$ candidates that do not pass the opposite charge requirement on the two leptons) which we refer to as the QCD control sample. The samples enriched in QCD all show the expected peaks at 0$^\circ$ and 180$^\circ$. 

When the $W$ recoil is less than 20 GeV the background is estimated by counting the excess of events in the distribution of $\Delta\phi(\rm{\ell-jet})$. The signal component is estimated by fitting a linear function to the middle part of the $\Delta\phi(\rm{\ell-jet})$ distribution. Almost all the $W$ candidates with recoil greater than 10 GeV come associated with at least one jet, and we account separately for events that do not have an associated 5 GeV jet. Since the $W$ candidates greatly outnumber the background events when the electron is isolated, the counting is done in bins of increasing isolation, and the background is extrapolated back to the signal region of $ISO_{0.25}<$ 1 GeV/$c$. The same background estimate is cross-checked by selecting events at high isolation ($6<ISO_{0.25}<$ 10 GeV/$c$) and using the fraction of isolated to non-isolated QCD events, seen in the QCD control sample, to predict the number in the signal region. Fig.~\ref{fig:qcd_typical}(b) shows the two-dimensional distribution of the recoil versus lepton isolation in the QCD control sample. 

We estimate 74$\pm$36 background events due to QCD in the 0$-$10 GeV recoil range and 30$\pm$11 in the 10$-$20 GeV recoil range. This includes an additional 10$\pm$7 events in the 0$-$10 GeV recoil bin due to $W$ events with no leading 5 GeV jet, as derived from the fraction of events  with and without a jet in the QCD control sample. The uncertainties include a systematic component due to the method. At higher $W$ recoil the estimate of the background is 3$\pm$2 events in both the 20$-$35 and 35$-$100 GeV bins. This is estimated with both the QCD control sample (by using the ratio of low to high recoil) and the direct counting of the excess of events at 0$^\circ$ and 180$^\circ$. In the latter, the non-uniform opening angle distribution of the recoiling jet and $W$-decay leptons is partially accounted for by a slope in the fit to the opening angle distribution. The small background contribution makes it unnecessary to accurately model the signal angular distribution.

\begin{figure}[p]
\begin{center}
\begin{tabular}{ll}
\epsfig{file=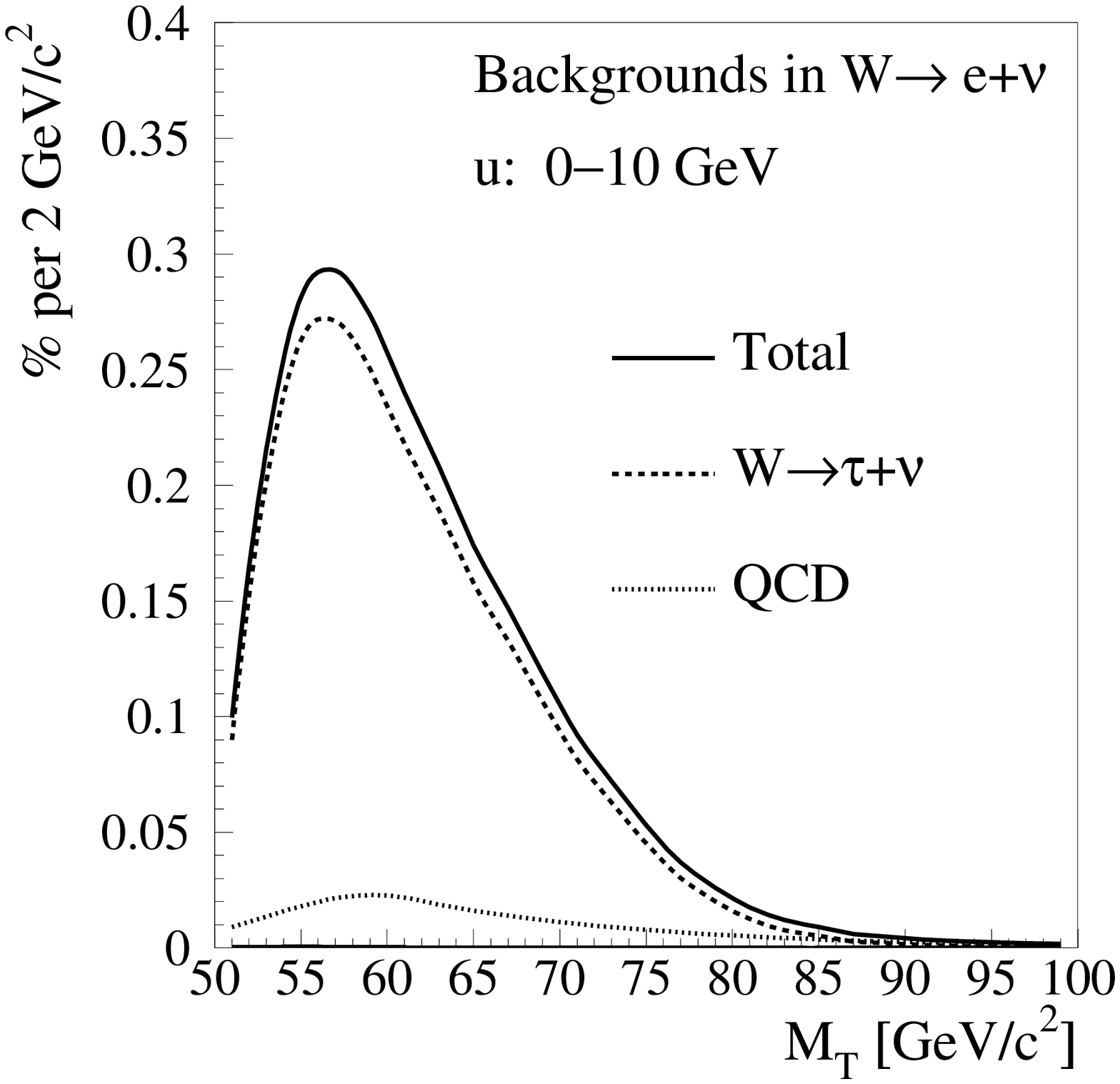,width=8cm,angle=0} &
\epsfig{file=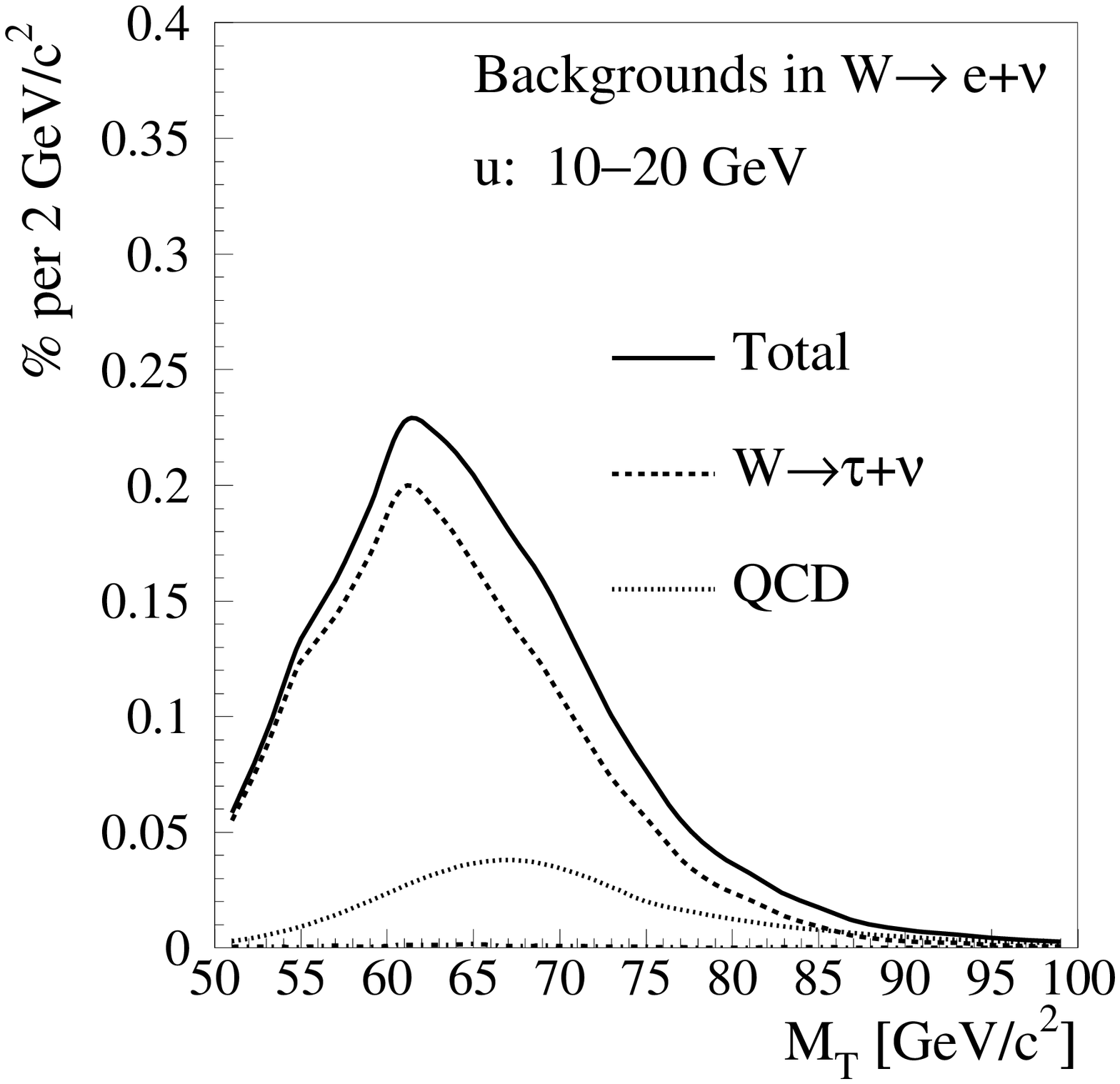,width=8cm,angle=0} \\
&\hspace{-5.2cm}\Large(a)\normalsize  \hspace{8.2cm}\Large(b)\normalsize \\
\epsfig{file=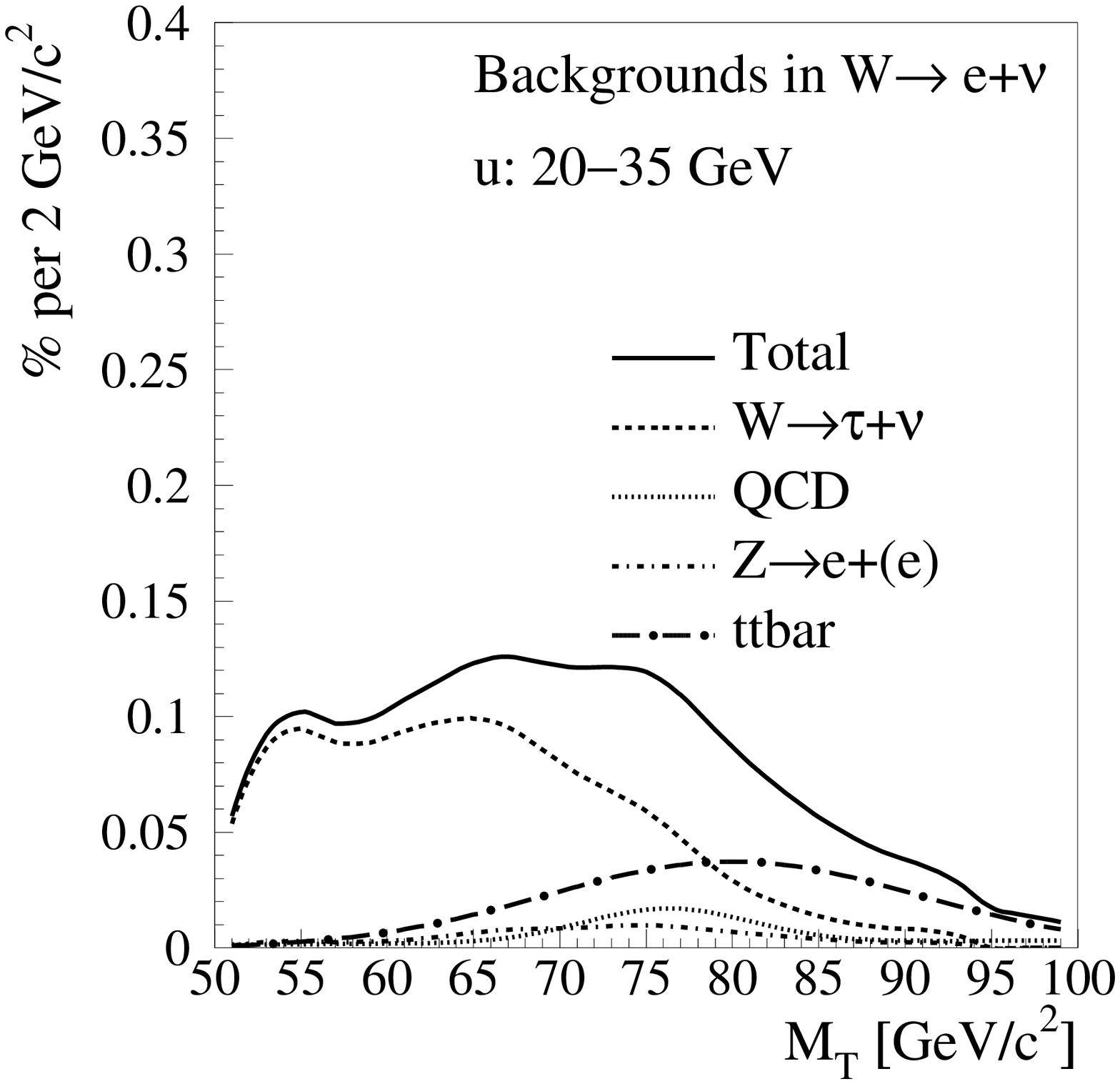,width=8cm,angle=0} & 
\epsfig{file=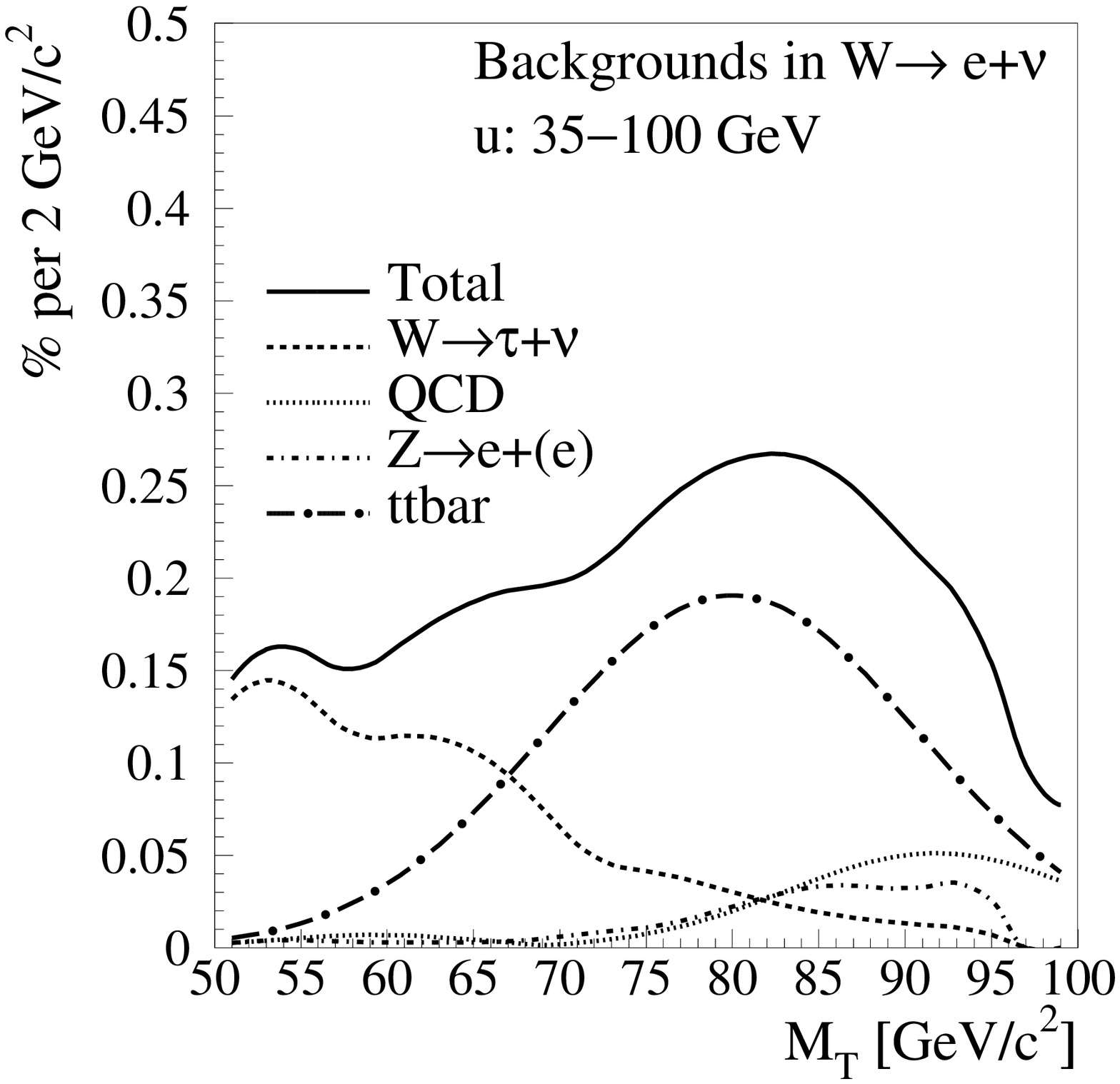,width=8cm,angle=0}  \\
&\hspace{-5.2cm}\Large(c)\normalsize  \hspace{8.2cm}\Large(d)\normalsize \\
\end{tabular}
\end{center}
\caption{\small{Electron channel: the transverse mass distribution from the background sources in four $W$ recoil ranges. The plots are in percentage of the $W$ data in the specific $p^W_T$ region.}}
\label{fig:total_ele_bkg_ptw}
\end{figure}

The shape of the transverse mass distribution of the QCD background is obtained by reversing the isolation cut and selecting events with anti-isolated electron tracks. The $M_T$ distribution shapes, at different recoil ranges, are seen to be largely independent of the anti-isolation cut. Fig.~\ref{fig:total_ele_bkg_ptw} shows the $M_T$ distribution of the backgrounds in the electron channel, scaled by the estimated amount as a percentage of the $W$ candidates. 

\begin{center}{\it 2: Muon channel}\end{center} 
QCD events can mimic W$\rightarrow\mu+\nu$ mainly in two ways. The first is when a heavy flavor quark in one of the jets decays into particles that include a high-$p_T$ muon (e.g. $b\rightarrow c+\mu+\nu$). In order for the muon and neutrino to have enough $p_T$ to pass the $W$ selection cuts, the $b$ quark needs to have a high transverse momentum, which leads to small opening angles. Therefore this type of event will have the muon and the neutrino almost parallel to one of the jets. The second major type of QCD background process occurs when a hadron is misidentified as a muon. The energy of one of the jets should also be incorrectly measured, in order to give the appearance of a high missing-$E_T$ signal. In this case, the neutrino and the muon will be reconstructed either nearly parallel to one jet or back-to-back and parallel to the two jets. Moreover, in both the processes considered, the muon is not likely to be isolated. 

\begin{figure}[p]
\begin{center}
\begin{tabular}{rl}
\epsfig{file=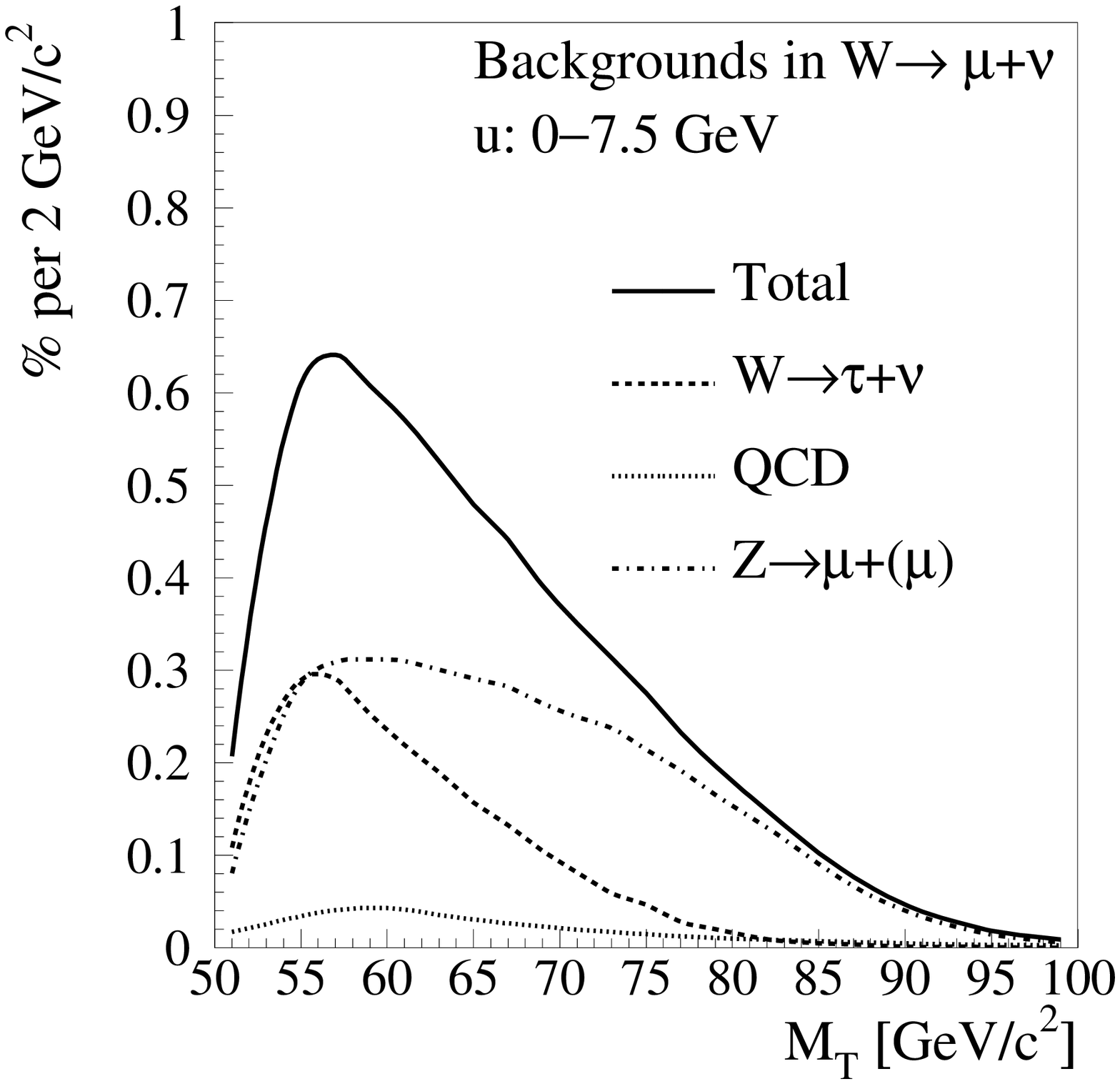,width=8cm,angle=0} &
\epsfig{file=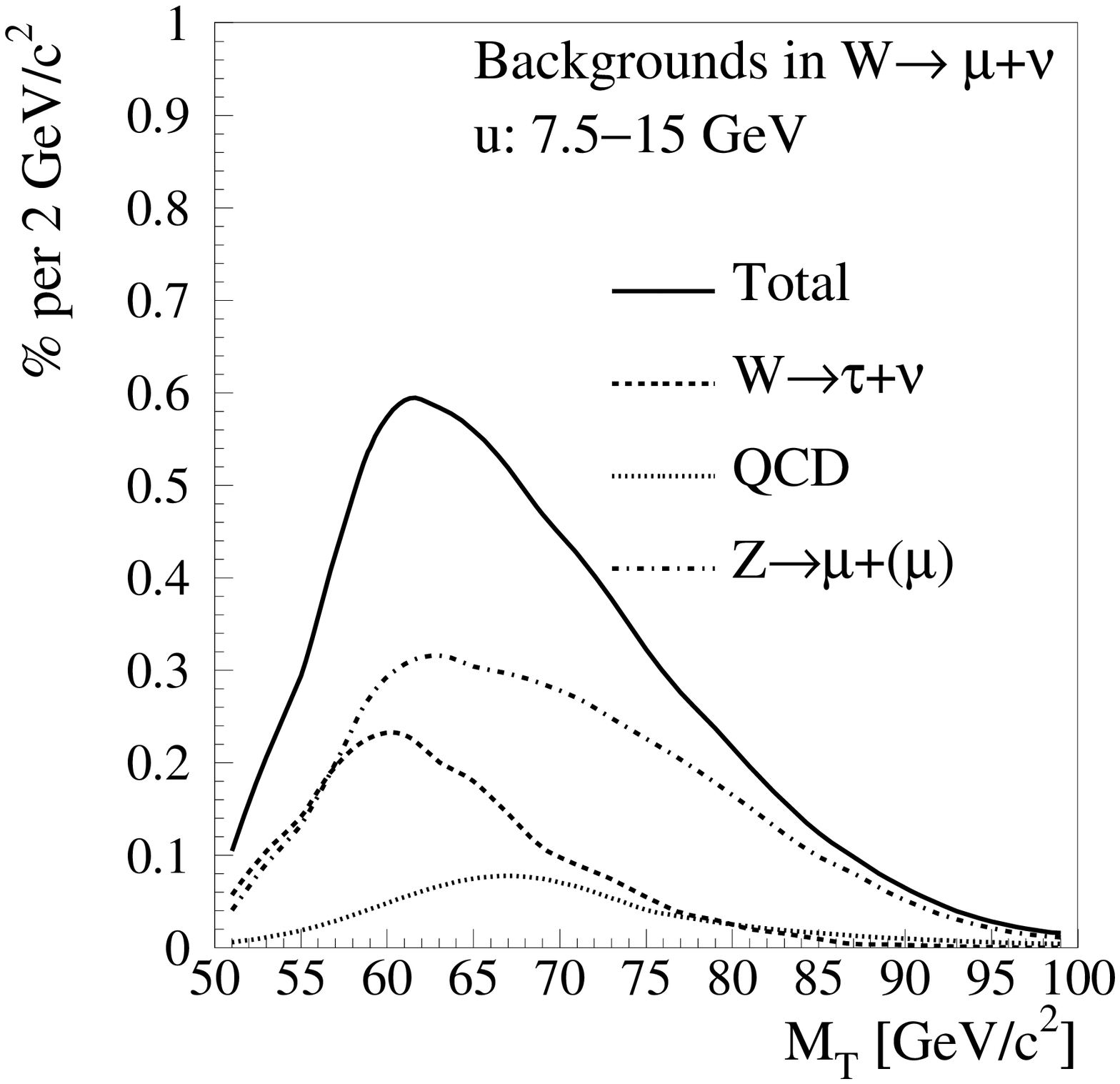,width=8cm,angle=0} \\
&\hspace{-5.2cm}\Large(a)\normalsize  \hspace{8.2cm}\Large(b)\normalsize \\
\epsfig{file=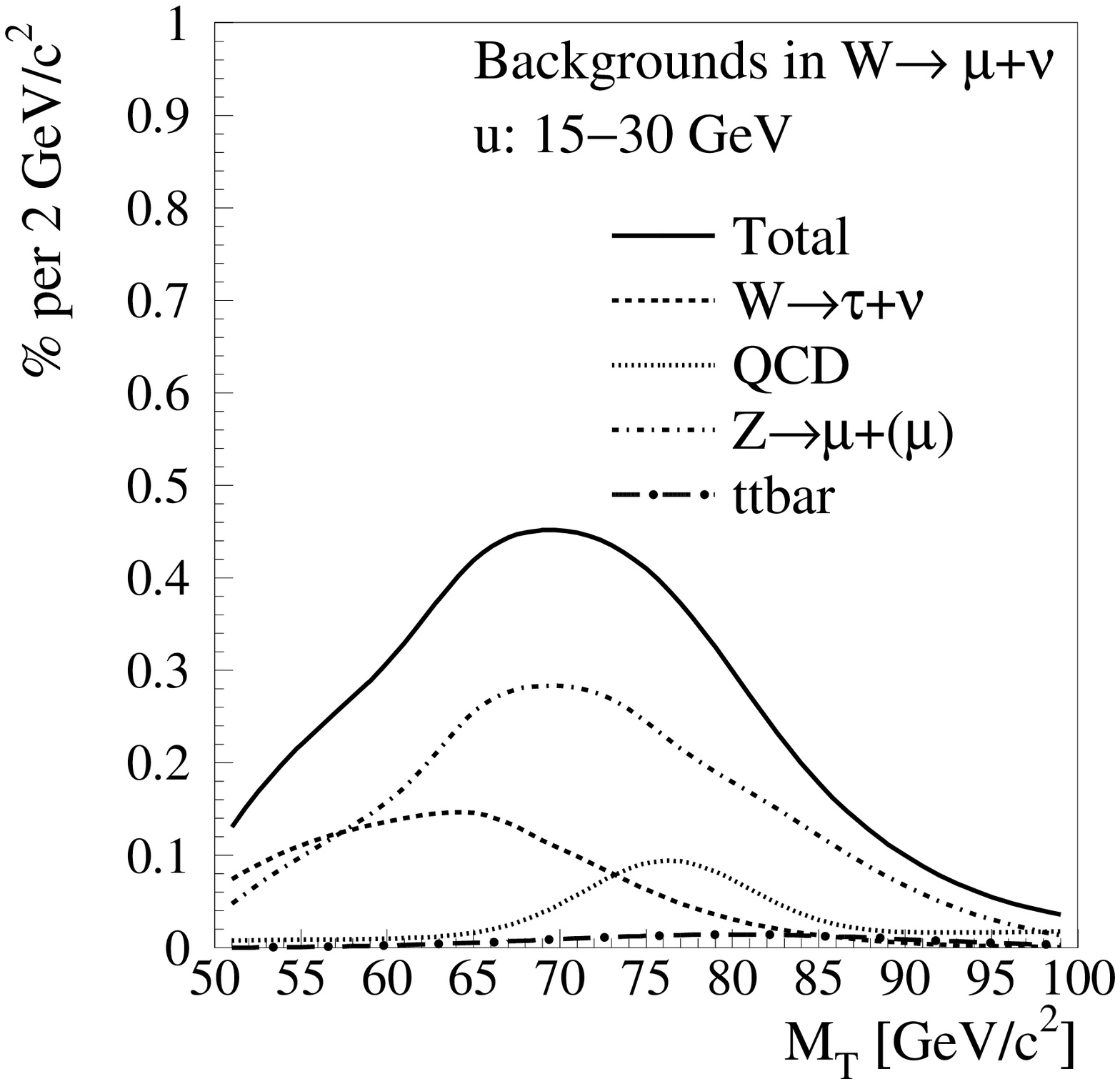,width=8cm,angle=0} & 
\epsfig{file=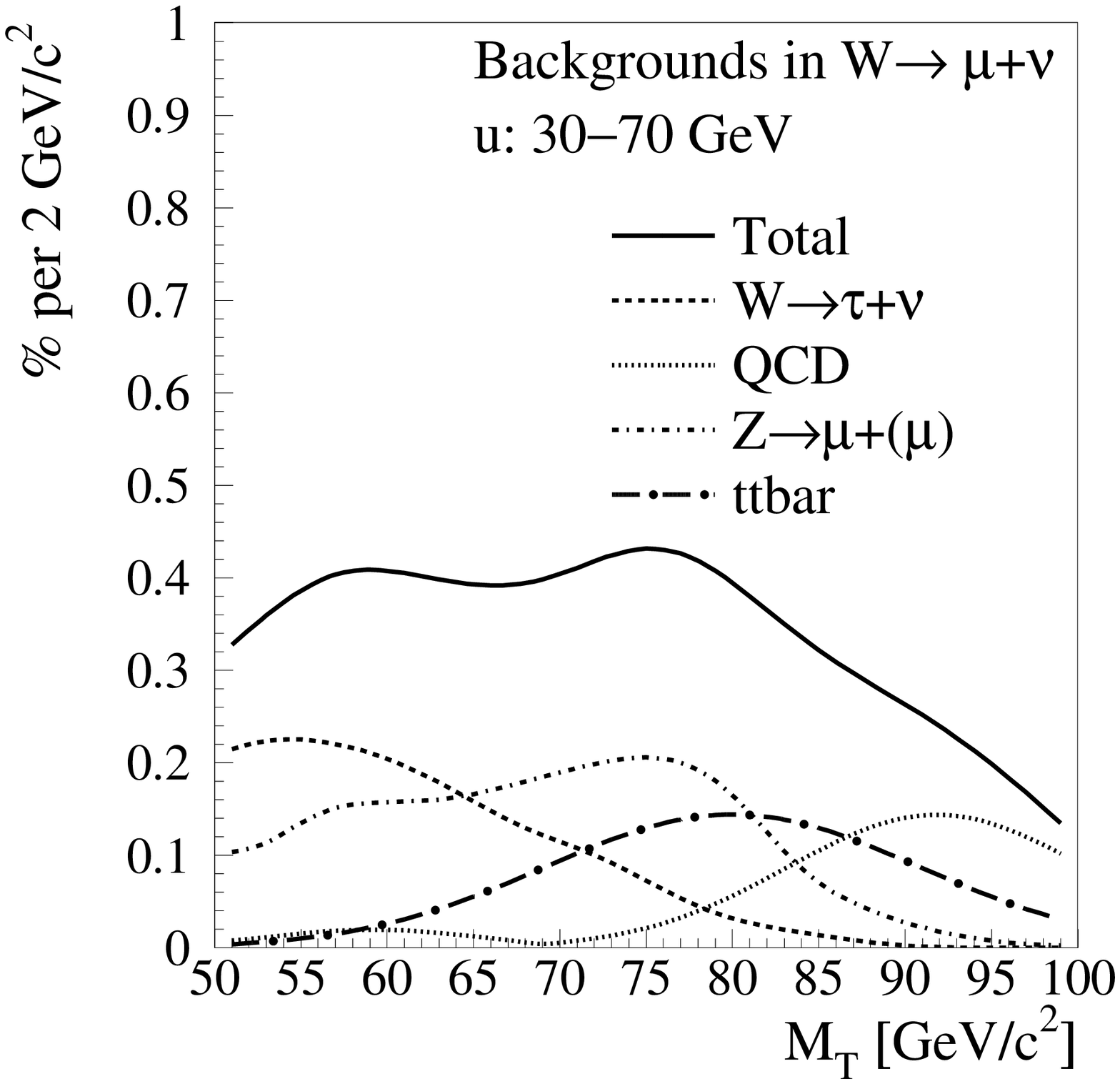,width=8cm,angle=0}  \\
&\hspace{-5.2cm}\Large(c)\normalsize  \hspace{8.2cm}\Large(d)\normalsize \\
\end{tabular}
\end{center}
\caption{\small{Muon channel: the transverse mass distribution from the background sources in four $W$ recoil ranges. The plots are in percentage of the $W$ data in the specific $p^W_T$ region.}}
\label{fig:total_muo_bkg_ptw}
\end{figure}

\begin{figure}[p]
\begin{center}
\begin{tabular}{rl}
\epsfig{file=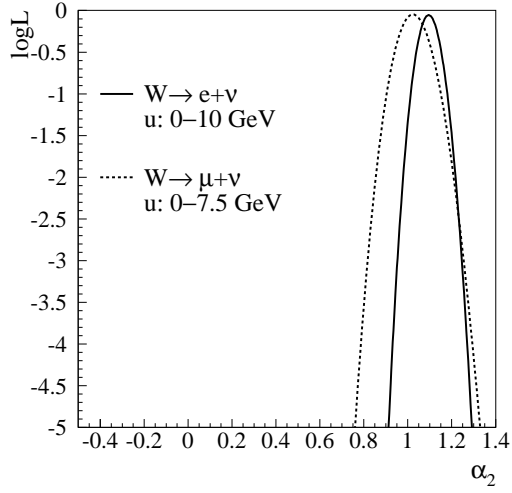,width=7.4cm,angle=0} &
\epsfig{file=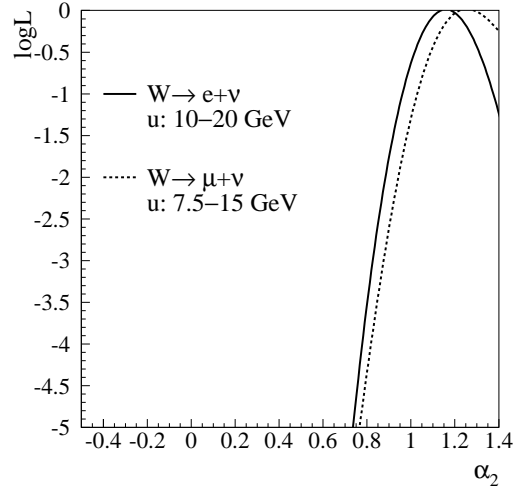,width=7.4cm,angle=0} \\
&\hspace{-4.2cm}\Large(a)\normalsize  \hspace{7.2cm}\Large(b)\normalsize \\
\epsfig{file=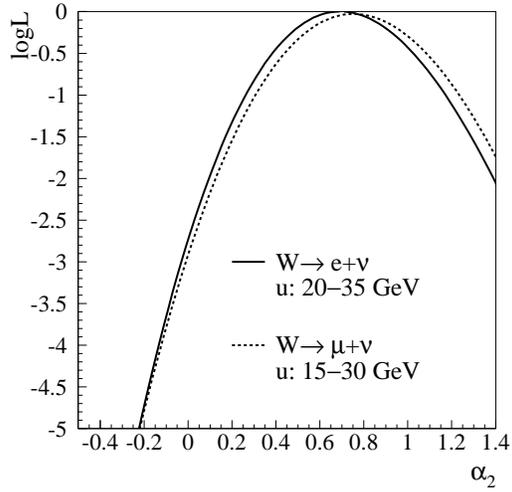,width=7.4cm,angle=0} &
\epsfig{file=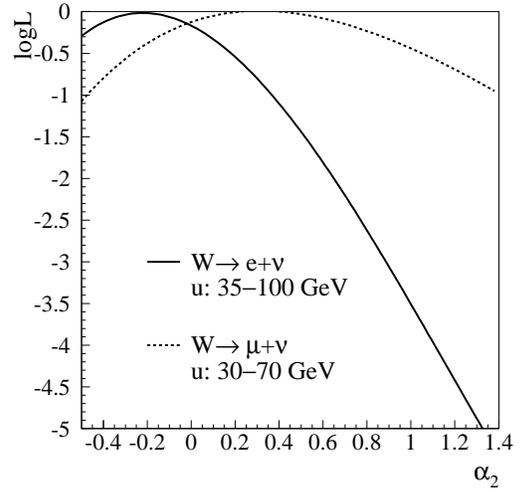,width=7.4cm,angle=0} \\
&\hspace{-4.2cm}\Large(c)\normalsize  \hspace{7.2cm}\Large(d)\normalsize \\
\end{tabular} 
\end{center}
\caption{\small{Likelihood functions of the fits for $\alpha_2$, in the four $W$-boson recoil regions for the electron and muon channels.}}
\label{fig:tmass_pt}
\end{figure}

\begin{figure}[p]
\begin{center}
\begin{tabular}{ll}
\epsfig{file=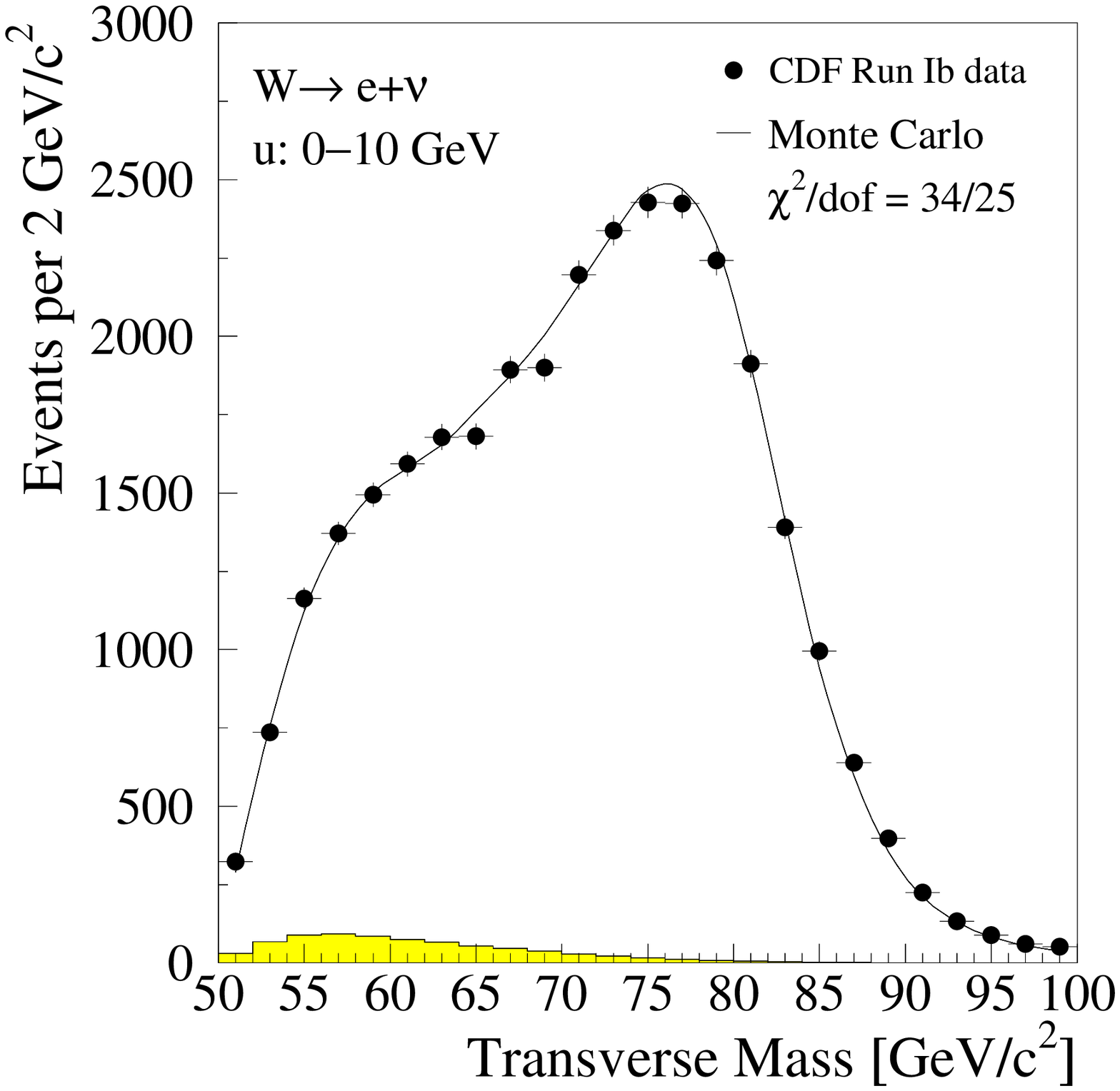,width=8cm,angle=0} &
\epsfig{file=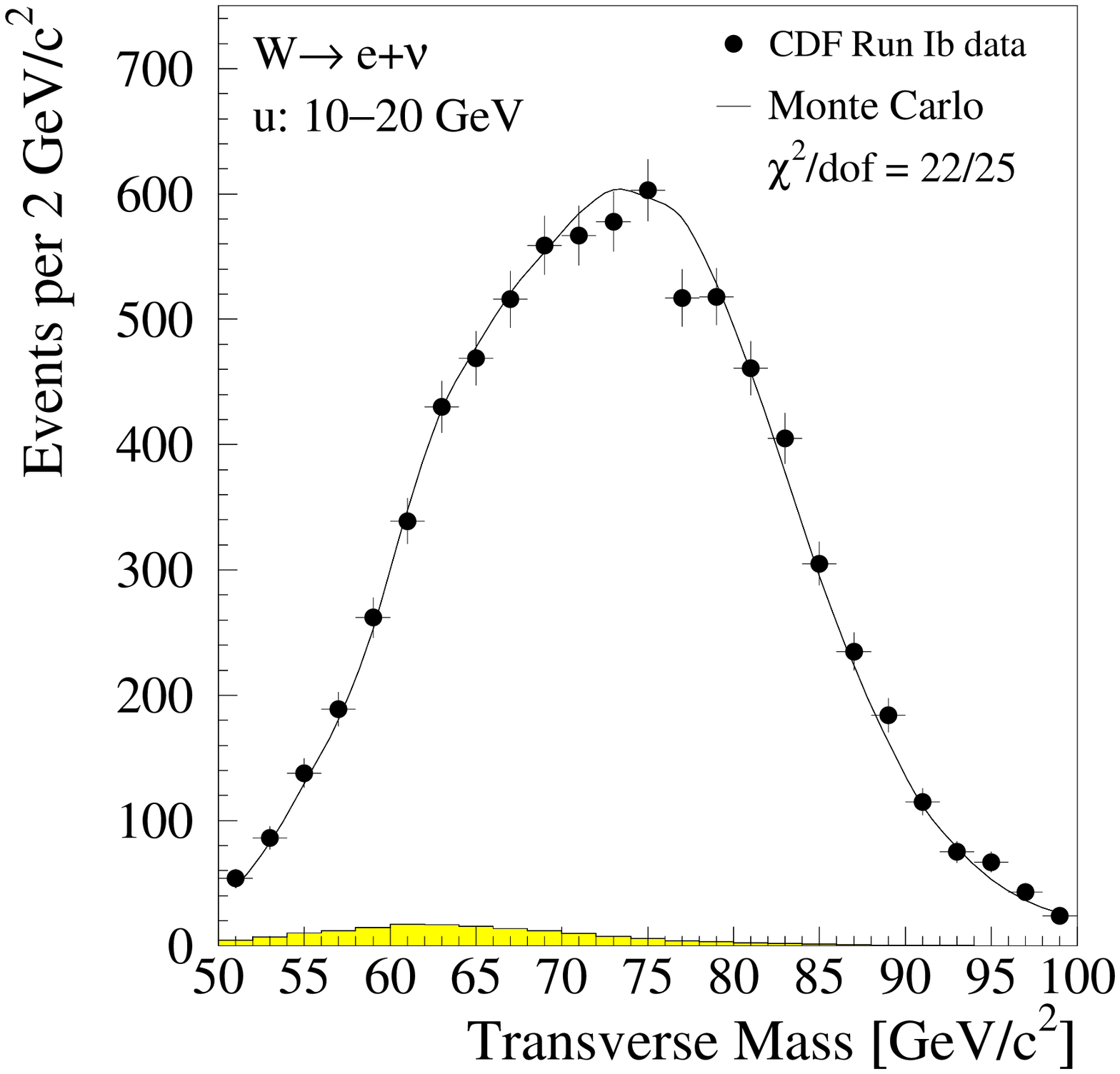,width=8cm,angle=0} \\
&\hspace{-4.2cm}\Large(a)\normalsize  \hspace{7.2cm}\Large(b)\normalsize \\
\epsfig{file=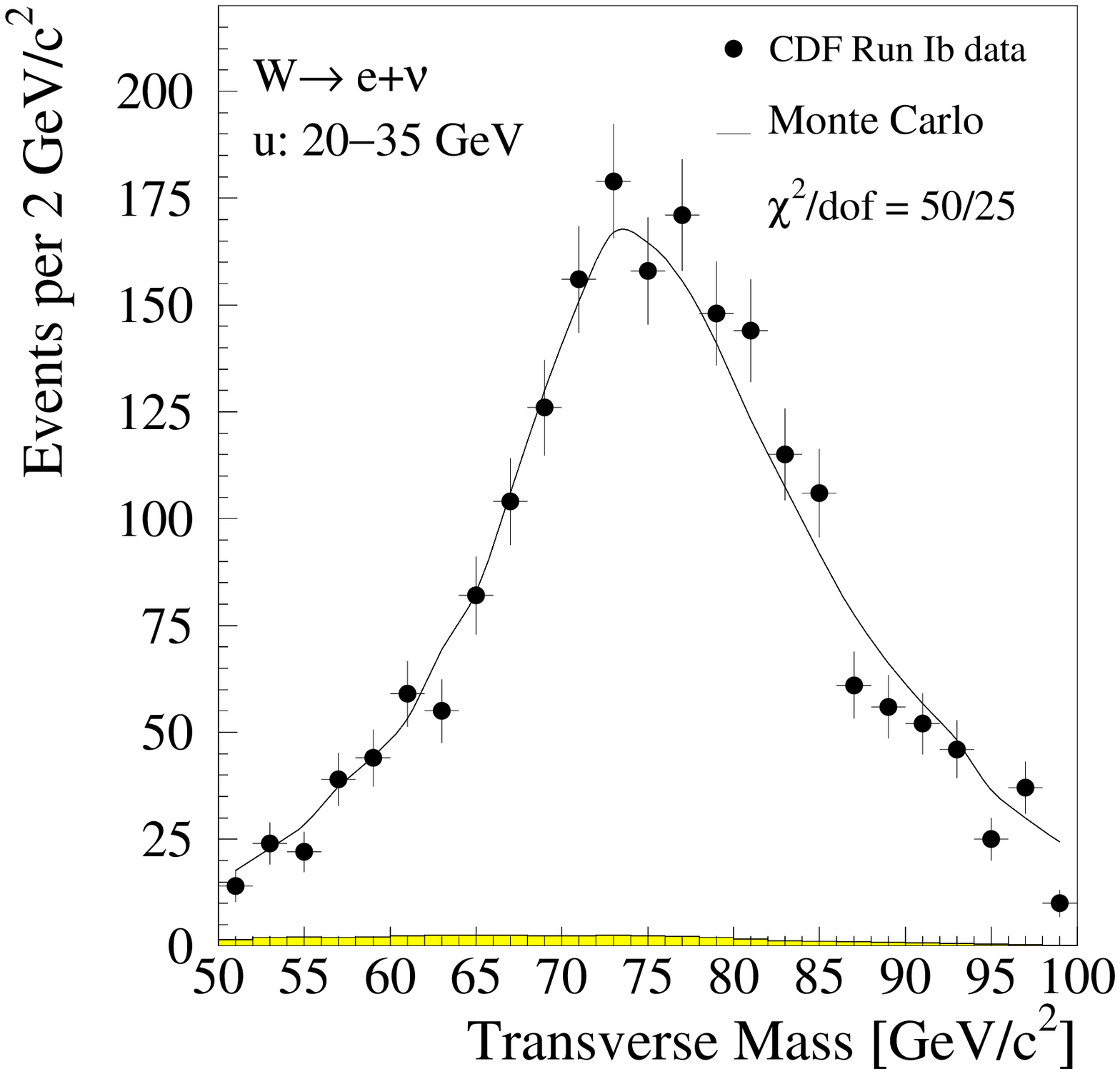,width=8cm,angle=0} &
\epsfig{file=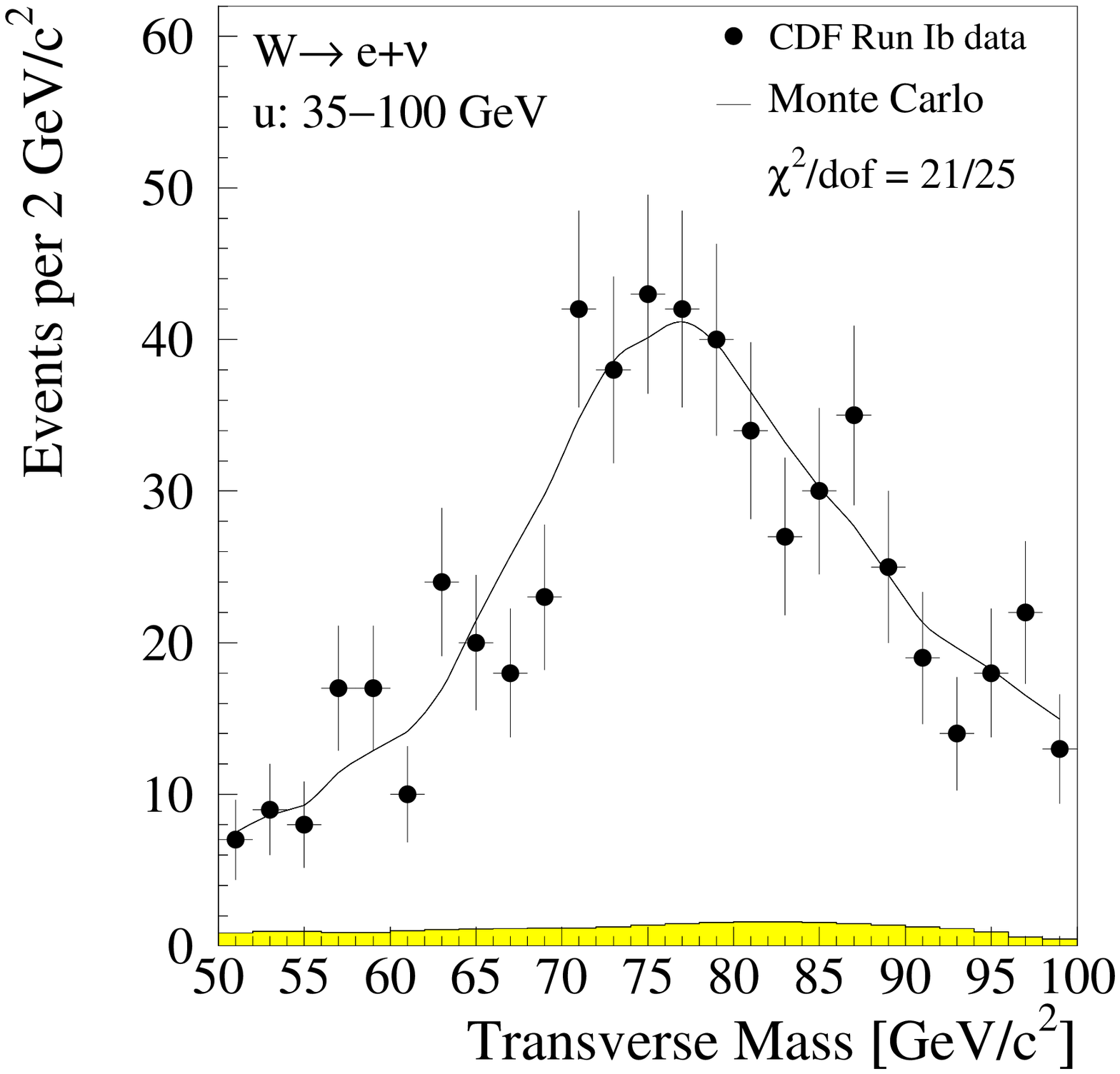,width=8cm,angle=0} \\
&\hspace{-4.2cm}\Large(c)\normalsize  \hspace{7.2cm}\Large(d)\normalsize \\
\end{tabular} 
\end{center}
\caption{\small{Comparison of the transverse mass distribution from the $W\rightarrow e+\nu$ data (filled circles) with the simulation (solid line) in four recoil regions. In the Monte Carlo simulation, $\alpha_2$ has been set to the best-fit value for each recoil range. The shaded histograms indicate the background contribution that is estimated to be present in the data and that has been added to the simulation.}}
\label{fig:ele_a2_bestfit_pt}
\end{figure}

\begin{figure}[p]
\begin{center}
\begin{tabular}{rl}
\epsfig{file=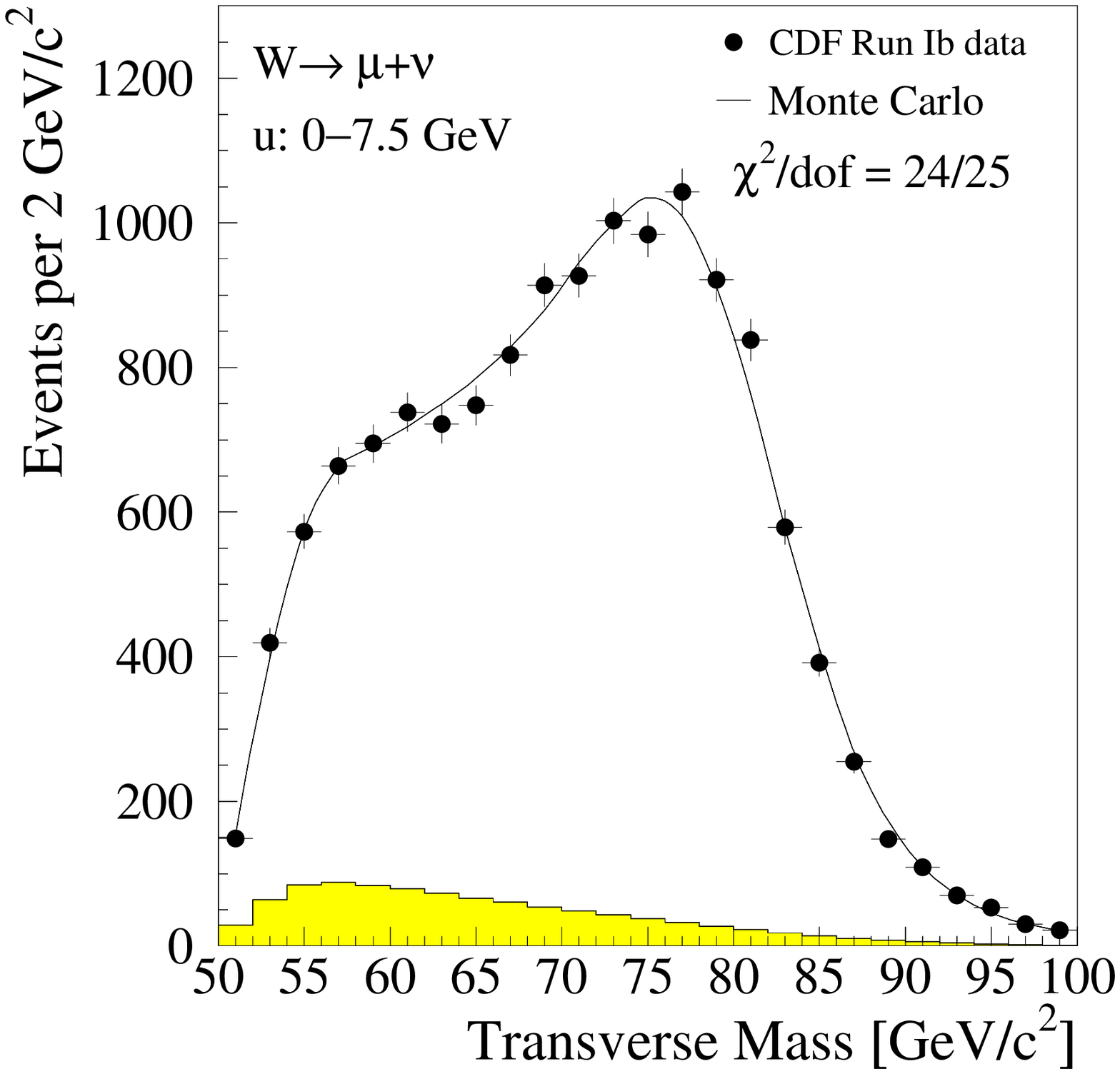,width=8cm,angle=0} &
\epsfig{file=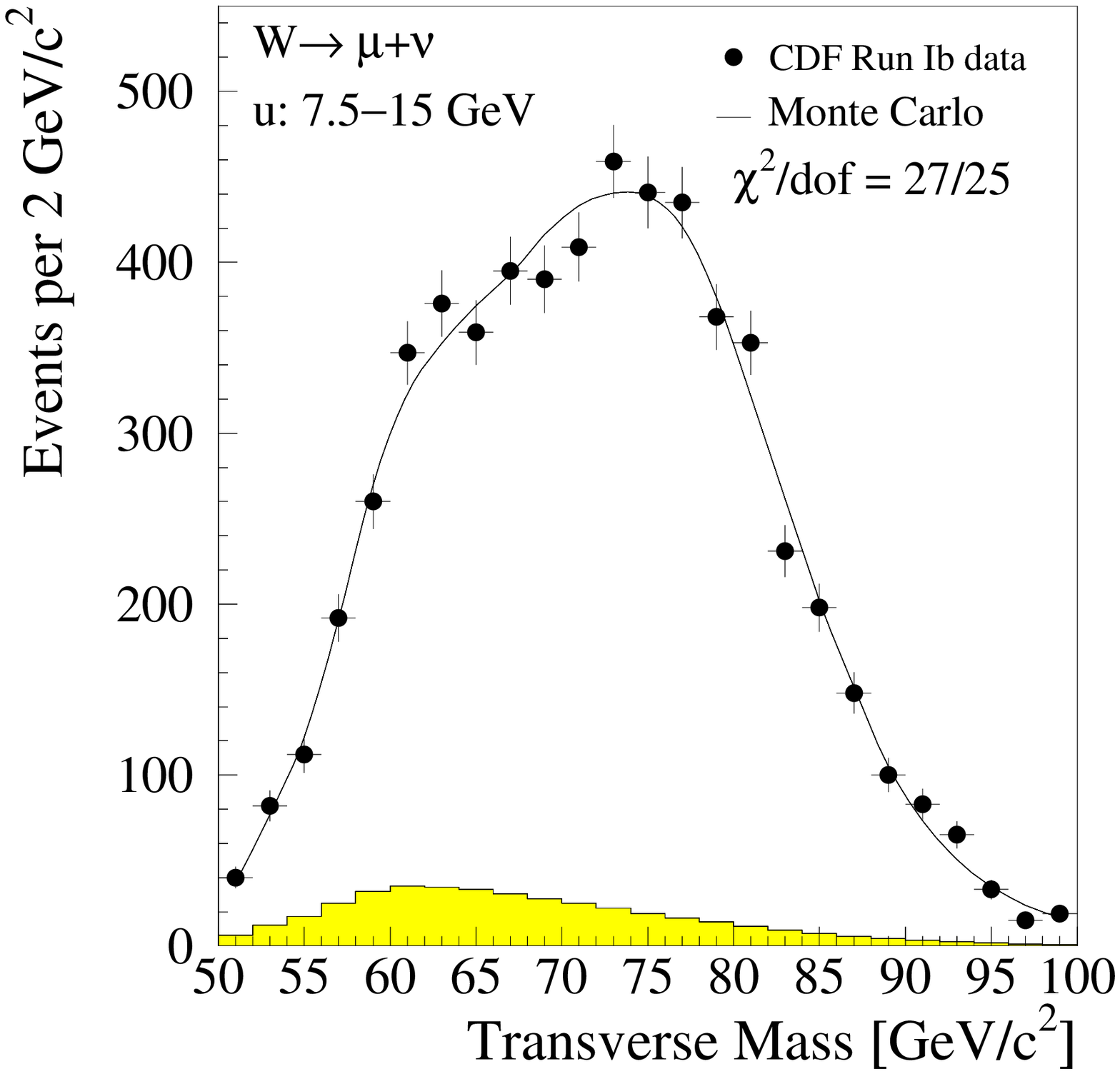,width=8cm,angle=0} \\
&\hspace{-4.2cm}\Large(a)\normalsize  \hspace{7.2cm}\Large(b)\normalsize \\
\epsfig{file=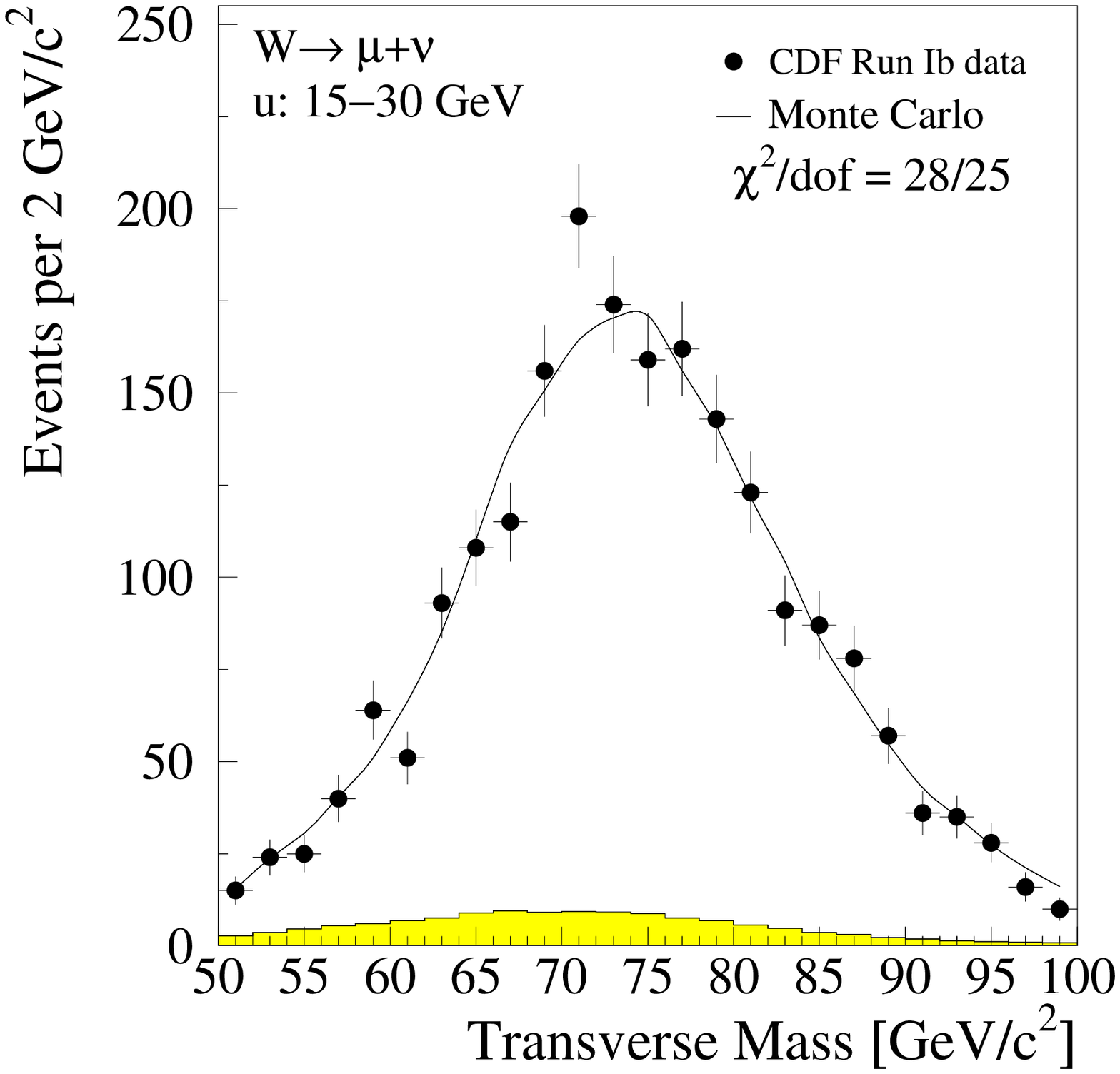,width=8cm,angle=0} &
\epsfig{file=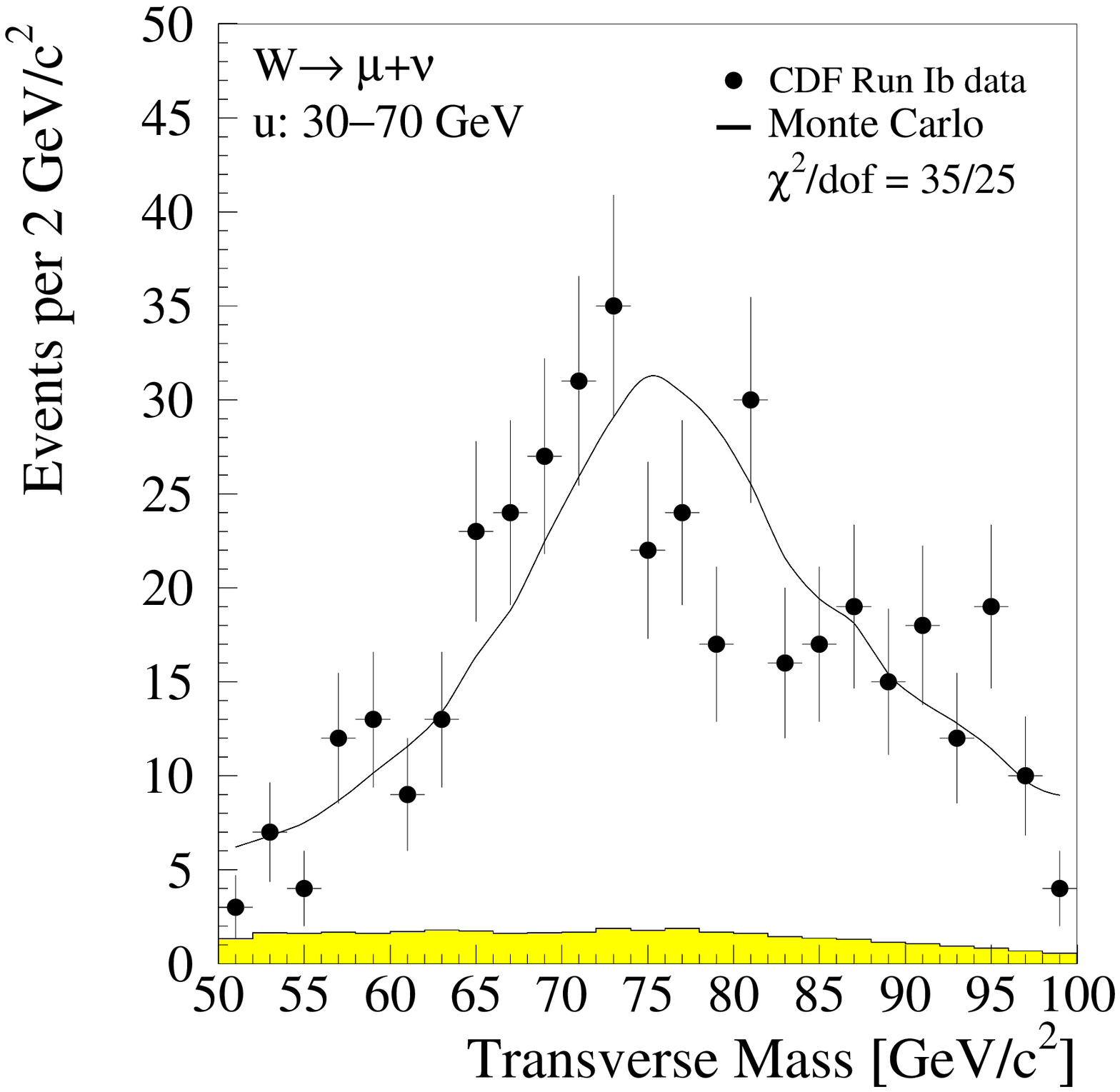,width=8cm,angle=0} \\
&\hspace{-4.2cm}\Large(c)\normalsize  \hspace{7.2cm}\Large(d)\normalsize \\
\end{tabular} 
\end{center}
\caption{\small{Comparison of the transverse mass distribution from the $W\rightarrow \mu+\nu$ data  (filled circles) with the simulation (solid line) in four recoil regions. In the Monte Carlo simulation, $\alpha_2$ has been set to the best-fit value for each recoil range. The shaded histograms indicate the background contribution that is estimated to be present in the data and that has been added to the simulation.}}
\label{fig:muo_a2_bestfit_pt}
\end{figure}

The QCD background to $W\rightarrow\mu+\nu$ events is estimated in the same way as for the electron channel in the four recoil bins. We expect $62\pm26$, 47$\pm$17, 17$\pm$11, and 6$\pm$5 events in the four recoil ranges. Fig.~\ref{fig:total_muo_bkg_ptw} shows the $M_T$ distribution of the backgrounds in the muon channel scaled by the estimated amount as a fraction (percent) of the $W$ candidates. 
\\
\begin{center}{\bf VII. FITS AND SYSTEMATIC UNCERTAINTIES}\end{center}
\begin{center}{\bf A: The likelihood fits}\end{center}
A set of Monte Carlo generated templates of the $M_T$ distribution is compared to the distribution derived from the data. When each template distribution is compared to the data, a likelihood is computed according to:
\begin{equation}
{\rm log} L(\alpha_2) =\sum_{i=1}^{\rm N_{bins}} n^{\rm data}_i ~{\rm log}\left[p_i^{\rm MC}(\alpha_2)\right],
\end{equation}
where the sum runs over the number of bins of the $M_T$ histogram, $n^{data}_i$ is the number of entries in each bin of the data histogram, and $p_i^{\rm MC}$ are the probabilities per-bin. The values of $p_i^{\rm MC}(\alpha_2)=n_i^{\rm MC}/n_{\rm tot}^{\rm MC}$ are given by the entries in the template histogram, one template for each value of $\alpha_2$. The maximum of the likelihood function locates the best estimate for the value of $\alpha_2$. Fig.~\ref{fig:tmass_pt} shows the likelihood functions in four different  $p_T^W$ regions for the electron and muon channels. The likelihood functions have been shifted vertically so that the maximum is always at zero. The 1$\sigma$ statistical uncertainty on each fit is evaluated at the points on the likelihood curve which are 1/2 unit below the maximum. The four recoil regions are 0$-$10, 10$-$20, 20$-$35, and 35$-$100 GeV/$c$ for the $W\rightarrow e+\nu$ data and 0$-$7.5, 7.5$-$15, 15$-$30, and 30$-$100 GeV/$c$ for the $W\rightarrow\mu+\nu$ data. The choice of the ranges is constrained by the sample size in the high-$p_T^W$ regions, due to the rapidly falling $p_T^W$ distribution. Moreover, the smaller sample of the muon channel is reflected in the recoil ranges covering lower $p_T^W$ values than in the electron channel.  Tables~\ref{tab:a2measel} and  \ref{tab:a2measmu} summarize the results of the fits for $\alpha_2$. Fig.~\ref{fig:ele_a2_bestfit_pt} and \ref{fig:muo_a2_bestfit_pt} show the transverse mass distribution of the data compared with the simulation, where $\alpha_2$ has been set to the best-estimate values.

\begin{center}{\bf B: Systematic uncertainties}\end{center}
Systematic uncertainties on the measurement of $\alpha_2$ for this analysis derive from the simulation of $W$ events, the detector response, and the estimate of backgrounds. Some of these, although classified as systematic, may be statistical in nature. This is the case for the detector recoil response and the $W$ transverse momentum spectrum, since they are derived from the $Z\rightarrow e^++e^-$ and $Z\rightarrow\mu^++\mu^-$ data samples. In the following, each source of systematic uncertainty is discussed and an estimate is determined for the shift on the measured values of $\alpha_2$. Tables~\ref{tab:a2measel} and \ref{tab:a2measmu} contain a summary of the various contributions and the total systematic uncertainty.

\begin{table}[h]
\begin{center}
\begin{tabular}{|l|r|r|r|r|}
\hline
Recoil range [GeV]      & 0$-$10       & 10$-$20      & 20$-$35      & 35$-$100  \\
\hline
$\alpha_2$ measured     & {\bf 1.09}   & {\bf 1.14}   & {\bf 0.67}   & {\bf $-$0.22}  \\
Statistical uncertainty & $\pm$0.05    & $\pm$0.13    & $\pm$0.29    & $\pm$0.36 \\
$\alpha_2$ predicted    & 0.98         & 0.84         & 0.55         & 0.25  \\
Mean $p_T^W$ [GeV/$c$]    & 6.2     & 15.9         & 33.3         & 59.2   \\
N$_{{\rm evt}}$              & 31363   & 7739         & 2033         & 595         \\
\hline
Systematic uncertainties:   $\qquad$ $\qquad$   &      &        &        &         \\
\hline
PDFs      &  $\qquad$ $\pm$0.01  & $\pm$0.01 & $\pm$0.01 & $\pm$0.01 \\
$W$ mass               & $\pm$0.02  & $\pm$0.01 & $\pm$0.04 & $\pm$0.04 \\
Input $p_T^Z$        & $\pm$0.02  & $\pm$0.03 & $\pm$0.03 & $\pm$0.04 \\
Recoil model         & $\pm$0.01  & $\pm$0.05 & $\pm$0.04 & $\pm$0.20 \\
Backgrounds       &$\pm$0.01   &$\pm$0.01  & $\pm$0.01 & $\pm$0.01\\
\hline
Combined systematic & $\pm$0.03 & $\pm$0.06 & $\pm$0.07 & $\pm$0.21  \\
\hline
\end{tabular}
\end{center}
\caption{\small {Summary of the measurement of $\alpha_2$ with the $W\rightarrow e+\nu$ data. The mean $p_T^W$ corresponding to each recoil range is the mean of the distribution of the ``true'' $W$ transverse momentum in the Monte Carlo simulation.}}  
\label{tab:a2measel}
\end{table}

\begin{table}[h]
\begin{center}
\begin{tabular}{|l|r|r|r|r|}
\hline
Recoil range [GeV]      &   0$-$7.5    &    7.5$-$15  &    15$-$30   &  30$-$70  \\
\hline
$\alpha_2$ measured     & {\bf 1.03}   & {\bf 1.24}   & {\bf 0.74}   & {\bf 0.24}    \\
Statistical uncertainty & $\pm$0.08    & $\pm$0.18    & $\pm$0.40    & $\pm$0.51 \\
$\alpha_2$ predicted    & 0.99         & 0.92         & 0.70         & 0.32  \\
Mean $p_T^W$ [GeV/$c$]    & 5.4          & 11.1         & 24.7         & 49.7   \\
N$_{{\rm evt}}$         & 13813        & 5910         & 2088         & 424    \\
\hline
Systematic uncertainties: $\qquad$ $\qquad$   &         &        &        &         \\
\hline
PDFs    &  $\qquad$ $\pm$0.01  & $\pm$0.01 & $\pm$0.01 & $\pm$0.01 \\
$W$ mass               & $\pm$0.02  & $\pm$0.01 & $\pm$0.04 & $\pm$0.04 \\
Input $p_T^Z$        & $\pm$0.02  & $\pm$0.03 & $\pm$0.03 & $\pm$0.04 \\
Recoil model         & $\pm$0.01  & $\pm$0.05 & $\pm$0.04 & $\pm$0.20 \\
Backgrounds       & $\pm$0.01  & $\pm$0.02 & $\pm$0.02 & $\pm$0.03 \\
\hline
Combined systematic &$\pm$0.03   &$\pm$0.06  & $\pm$0.07 & $\pm$0.21 \\
\hline
\end{tabular}
\end{center}
\caption{\small{Summary of the measurement of $\alpha_2$ with the $W\rightarrow \mu+\nu$ data. The mean $p_T^W$ corresponding to each recoil range is the mean of the distribution of  the ``true'' $W$ transverse momentum in the Monte Carlo simulation.}}  
\label{tab:a2measmu}
\end{table}

\begin{center}{\it 1. Event Selection Bias}\end{center}
The electron isolation requirement may introduce a bias on the measurement of $\alpha_2$. For example, if the electron travels close to the recoil, there is greater opportunity for the event to be rejected. Also, there could be a correlation of the selected sample with $\alpha_2$, which is correlated with the QCD activity in the event. This bias is investigated by removing the isolation requirement, evaluating appropriately the increase in backgrounds, and measuring the change in $\alpha_2$. The maximum shifts observed are within the systematic uncertainty of the background determination. Moreover, by changing widely $\alpha_2$ in the simulation, the spectrum of the opening angle between recoil and electron directions is not significantly affected. We do not apply an isolation requirement to the muon channel.

\begin{center}{\it 2. Parton Density Functions}\end{center}
The parton distribution functions are used in the Monte Carlo simulation to determine the quark content of the proton, and hence the rapidity distribution of the generated $W$ bosons. The set of PDFs used to simulate the events in this analysis is MRS-R2 \cite{mrs}. These PDFs describe well the CDF low-$\eta$ $W$-charge asymmetry data. To evaluate the impact of the choice of PDFs on the measurement of $\alpha_2$, two Monte Carlo samples have been generated with MRMS-D$^-$ and MRMS-D0, sets that were not tuned on CDF data and differ significantly from MRS-R2. $\alpha_2$ has been measured with both sets of PDFs. The observed shifts are $\pm0.01$ in all recoil regions, a small fraction of the statistical uncertainty. This is conservatively taken to be the systematic uncertainty due to the choice of PDFs.

\begin{center}{\it 3. The $W$ mass value}\end{center}
The transverse mass distribution is sensitive to the value of the $W$ mass used in the Monte Carlo simulation. The dependence comes from the fact that the transverse mass spectrum has a Jacobian peak at about the value of the $W$ mass. The value of the $W$ mass in the Monte Carlo simulation is set to the LEP average \cite{lep2mass} 80.412$\pm$0.042 MeV/$c^2$, in order to be independent of the value measured at CDF. An uncertainty on $M_W$ of 40 MeV$/c^2$ corresponds to a systematic uncertainty on $\alpha_2$ of 0.01$-$0.04.

\begin{center}{\it 4. $p_T^W$ spectrum}\end{center}
The $W$ transverse momentum spectrum is derived from the $Z$ sample by measuring $p_T^Z$, and using the relatively well known ratio of the $p_T^W/p_T^Z$ distributions from theory. The $p_T^Z$ distribution is measured from both the $Z\rightarrow e^++e^-$ and $Z\rightarrow\mu^++\mu^-$ data, and then averaged. To account for statistical and systematic uncertainties in determining the $p_T^Z$ spectrum, additional MC datasets are generated using the $p_T^Z$ from the electron or the muon $Z$-decay channels only. The measured $\alpha_2$ shifts by between 0.02 and 0.04.

\begin{center}{\it 5. Recoil Model}\end{center}
The recoil model consists of response and resolution functions derived from the $Z\rightarrow e^++e^-$ and $Z\rightarrow \mu^++\mu^-$ data. There are statistical uncertainties in the coefficients of the model, which are used here to evaluate a systematic uncertainty. Each of the parameters is changed and the $\alpha_2$ value is measured. The dispersion of the set of new measurements is taken as the systematic uncertainty, which increases with $p_T^W$ as shown in Tables~\ref{tab:a2measel} and \ref{tab:a2measmu}. The recoil model is one of the main sources of uncertainty here since it is constrained with a statistical sample much smaller than the $W$ sample itself. The impact of a slight disagreement between the $W$ recoil distribution in data and simulation has been estimated by shifting the edges of the recoil ranges one at a time by 0.1 GeV/$c$, only in the data but not in the Monte Carlo simulation, simulating event migration between bins. The value of 0.1 GeV/$c$ is the difference between the mean of the recoil distributions in the data and in the MC. The coefficient $\alpha_2$ has been observed to shift between 0.01 and 0.04 in the four bins. This is included in the quoted systematic uncertainty due to the recoil model.

\begin{center}{\it 6. The angular coefficients and $\alpha_1$ input value}\end{center}
Although the distribution of $|\cos\theta_{CS}|$, and hence $M_T$, should only depend on $\alpha_2$ and all the remaining angular coefficients should integrate out in practice geometric acceptance causes some angular coefficients to have a residual effect on the shape of the $M_T$ distribution. Coefficients A$_1$, A$_5$, A$_6$, A$_7$ are predicted to be negligible in the Standard Model and are set to zero. A$_2$ and A$_3$ are kept in the angular distribution (see Eq.~(\ref{equ:qcdangular})) and are set to their Standard Model values. As an estimate of the sensitivity to these terms, neglecting $A_2$ and $A_3$ results in a shift in the value of $\alpha_2$ of 0.02$-$0.07 in the four $p_T^W$ bins. These values are not included in the systematic uncertainty since the uncertainty on the theoretical SM calculation is expected to be much smaller than 100\%.

The coefficient $\alpha_1$ also enters the $M_T$ distribution. However, when fitting for $\alpha_2$ at low $p_T^W$, $\alpha_1$ cannot be set to the SM expected value, due to the requirement of positive event weights expressed in Eq.~(\ref{equ:weight_boundaries}). $\alpha_1$ is therefore set to $2\sqrt{\alpha_2}$, which lies in vicinity of the SM path for low $p_T^W$. With this choice, Eq.~(\ref{equ:weight_boundaries}) translates into a condition for $(1\pm\sqrt{\alpha_2} \cos\theta_{CS})^2$, which is always positive and prevents assigning negative weights in the region around the Quark Parton Model point. A negligible change in the measured $\alpha_2$ is visible by setting $\alpha_1(p_T^W)$ to different paths around the SM expectation. For higher $p_T^W$ ($\geq 20$ GeV/$c$), $\alpha_1$ is set to the full SM prediction as there is no danger of assigning negative weights in that region.

\begin{center}{\it 7. Backgrounds}\end{center}
The main sources of uncertainty due to backgrounds come from the estimates of the QCD and $t\bar{t}$ contributions. The QCD background is estimated from the data using the lepton isolation and the angular distribution between the lepton and the jets in the event and the $t\bar{t}$ background is taken from \cite{ttbarref}. The systematic uncertainty on the measured values of $\alpha_2$ is derived  by changing the QCD and $t\bar{t}$ background contents in each $p_T^W$ range by the uncertainty given in the background estimate results in Tables~\ref{tab:backelec} and \ref{tab:backmuon}. A maximum shift of 0.03 on $\alpha_2$ is observed.

\begin{figure}[p]
\begin{center}
\epsfig{file=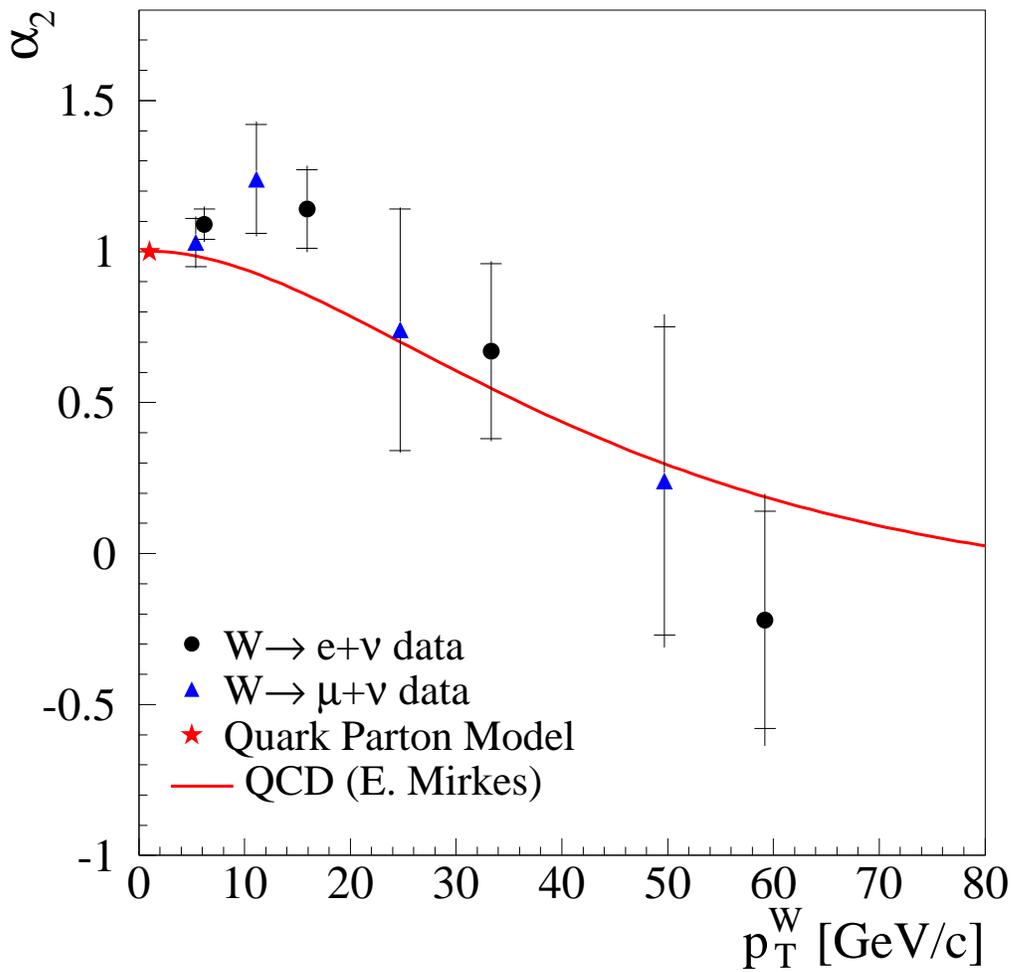,width=15cm,angle=0}
\end{center}
\caption{\small{Measurement of $\alpha_2$ with the electron (filled circles) and the muon (triangles) channels. The error bars include statistical and systematic uncertainties, and the tick marks show statistics only.}}
\label{fig:alpha2_all}
\end{figure}

\begin{center}{\bf VIII. RESULTS AND CONCLUSIONS}\end{center}
 Fig.~\ref{fig:alpha2_all} shows the results of this measurement on a plot of $\alpha_2$ versus $p_T^W$. The position of the points on the $x$ axis has been determined by using the mean of the Monte Carlo distribution of $p_T^W$ corresponding to each recoil range. The solid curve represents the Standard Model prediction reported in \cite{mirkes}. The trend is a decrease of $\alpha_2$ with increasing $p_T^W$, which corresponds to an increase of the longitudinal component of the $W$ polarization. The rate measured from a linear fit is $\sim$15\% per 10 GeV/$c$. The four measurement points from the electron channel can be used together with those from the muon channel to compute a $\chi^2$ with respect to the Standard Model expectation. The result is $\chi^2$=1.5, normalized for 8 degrees of freedom and considering statistical and systematic uncertainties. 

\begin{figure}[p]
\begin{center}
\epsfig{file=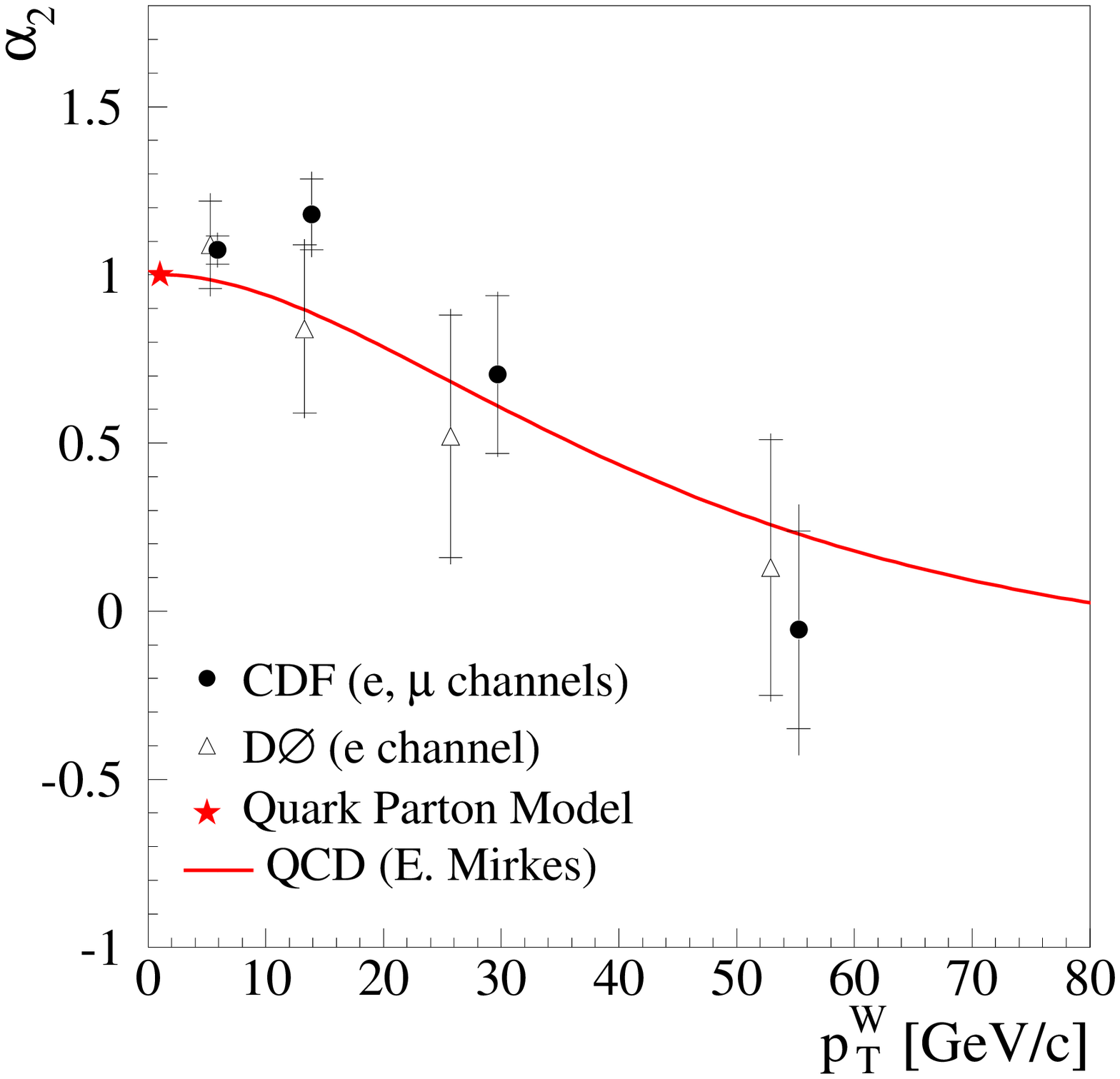,width=15cm,angle=0}
\end{center}
\caption{\small{Measurement of $\alpha_2$ combining the electron and the muon channels (filled circles). The error bars show the combination of the statistical and systematic uncertainties. The D\O~measurement (open triangles) is from \cite{d0paper}.}}
\label{fig:combine}
\end{figure}

The measurements of $\alpha_2$ with the electron and muon channels are combined in Fig.~\ref{fig:combine} and Table~\ref{tab:a2combined}. The position in $p_T^W$ is determined by a weighted mean of each pair of electron and muon measurements. The values of $\alpha_2$ are then scaled at the common $p_T^W$ value using a linear fit and then averaged taking into account the size of the statistical uncertainties. Systematic uncertainties are completely correlated between the electron and muon channels. The triangles are from \cite{d0paper} and represent the current best values.  
\begin{table}[h]
\begin{center}
\begin{tabular}{|l|c|c|}
\hline
$p_T^W$ [GeV/$c$] & $\alpha_2$ (CDF combined) & $\alpha_2$ (theory)   \\
\hline
 5.9             & 1.07$\pm$0.04(stat)$\pm$0.03(syst) & 0.98   \\
13.9             & 1.18$\pm$0.10(stat)$\pm$0.06(syst) & 0.89   \\
29.7             & 0.70$\pm$0.23(stat)$\pm$0.07(syst) & 0.61   \\
55.3             & -0.05$\pm$0.29(stat)$\pm$0.21(syst) & 0.23   \\
\hline
\end{tabular}
\end{center}
\caption{\small {Summary of the measured $\alpha_2$ combining the electron and muon channels.}}  
\label{tab:a2combined}
\end{table}

In conclusion, we have measured the leptonic polar-angle distribution coefficient $\alpha_2$ as a function of the transverse momentum of the $W$ boson. The results obtained from the electron and muon channels are combined together and the measurement reduces by about 50\% the uncertainty on the current best values up to $p_T^W\sim 30$ GeV/$c$. The result is in good agreement with the Standard Model expectation up to NLO, whereby $\alpha_2$ decreases with $p_T^W$ as a consequence of QCD corrections to the $W$ polarization. Since the uncertainty is largely dominated by statistics especially at higher $W$ transverse momenta, this measurement can significantly benefit from the larger data sample of $p\bar{p}$ collisions at $\sqrt{s}=1.96$ TeV that is being collected at CDF in Run II. 

\begin{center}{\bf ACKNOWLEDGMENTS}\end{center}
We thank the Fermilab staff and the technical staffs of the participating institutions for their vital contributions.  
This work was supported by the U.S. Department of Energy and National Science Foundation; the Italian Istituto Nazionale di Fisica Nucleare; the Ministry of Education, Culture, Sports, Science, and Technology of Japan; the Natural Sciences and Engineering Research Council of Canada; the National Science Council of the Republic of China; the Swiss National Science Foundation; the A.P. Sloan Foundation; the Bundesministerium fuer Bildung und Forschung, Germany; and the Korea Science and Engineering Foundation (KoSEF); the Korea Research Foundation;  the Comision Interministerial de Ciencia y Tecnologia, Spain; and the Council for the Central Laboratory of the Research Councils, UK.

%%%%%%%%%%%%%%%Bibliography


\begin{thebibliography}{99}
\bibitem{cesshape} The CES determines the transverse development of an electromagnetic shower at shower maximum by measurement of the charge deposition on orthogonal strips and wires. The profile of the CES strip or wire cluster is compared to a template and a $\chi^2$ is formed as a goodness of the fit.
\end{thebibliography}
\end{document}